\documentclass[conference]{IEEEtran}

\usepackage[]{hyperref}
\usepackage{algorithm}
\usepackage{listings}
\usepackage[noend]{algpseudocode}
\usepackage{graphicx}
\usepackage{textcomp}
\usepackage[dvipsnames]{xcolor}
\usepackage{booktabs}
\usepackage{enumitem}
\usepackage{comment}
\usepackage{amsmath}
\usepackage{setspace}

\graphicspath{{figures/}}

\newcommand{\rvv}{RISC-V V}
\newcommand{\mte}{MTE}
\newcommand{\nexperiments}{75}

\newcommand{\vlong}{Vector 1KB}
\newcommand{\vlonger}{Vector 2KB}
\newcommand{\amx}{AMX}
\newcommand{\sifive}{SiFiveInt}

\newcommand{\sv}{Vector 1KB}
\newcommand{\slv}{Vector 2KB}
\newcommand{\ssf}{SiFiveInt}
\newcommand{\samx}{$MTE_{8s}$}
\newcommand{\smtev}{$MTE_{32v}$}
\newcommand{\smtes}{$MTE_{32s}$}

\newcommand{\celvv}{78.7\%}

\newcommand{\celvvi}{70.7\%}

\newcommand{\telvv}{32.7\%}

\newcommand{\telvvi}{19.5\%}

\newcommand{\xevi}{6.3\%}

\newcommand{\xesfi}{11.3\%}

\newcommand{\xevmtei}{29.1\%}
\newcommand{\xesmtei}{40.3\%}

\newcommand{\xesfii}{26.8\%}

\newcommand{\xevmteii}{51.8\%}
\newcommand{\xesmteii}{67.3\%}

\newcommand{\xesfiii}{36.6\%}

\newcommand{\xeviv}{53.1\%}

\newcommand{\xesfiv}{42.9\%}

\newcommand{\xesfv}{41.8\%}

\newcommand{\xesfvi}{43.2\%}

\newcommand{\xevmtevi}{86.8\%}
\newcommand{\xesmtevi}{93.2\%}
\newcommand{\suv}{2.3$\times$}
\newcommand{\sulv}{2.11$\times$}
\newcommand{\susf}{1.98$\times$}
\newcommand{\susamx}{1.16$\times$}
\newcommand{\ssuv}{2.67$\times$}
\newcommand{\ssulv}{2.45$\times$}
\newcommand{\ssusf}{2.3$\times$}
\newcommand{\ssusamx}{1.35$\times$}
\newcommand{\supmtev}{\susamx{}} 
\newcommand{\supmtes}{\ssusamx{}} 

\AtBeginDocument{%
  \providecommand\BibTeX{{%
    Bib\TeX}}}

\def\BibTeX{{\rm B\kern-.05em{\sc i\kern-.025em b}\kern-.08em
    T\kern-.1667em\lower.7ex\hbox{E}\kern-.125emX}}

\begin{document}
\pagestyle{plain}

\title{A Flexible Instruction Set Architecture\\for Efficient GEMMs}

\author{\IEEEauthorblockN{Alexandre de Limas Santana}
\IEEEauthorblockA{Barcelona Supercomputing Center\\
Universitat Politècnica de Catalunya\\
Email: alexandre.delimassantana@bsc.es}
\and
\IEEEauthorblockN{Adrià Armejach Sanosa}
\IEEEauthorblockA{Barcelona Supercomputing Center\\
Universitat Politècnica de Catalunya\\
Email: adria.armejach@bsc.es}
\and
\IEEEauthorblockN{Francesc Martinez}
\IEEEauthorblockA{Barcelona Supercomputing Center\\
Universitat Politècnica de Catalunya\\
Email: francesc.martinez@bsc.es}
\and
\IEEEauthorblockN{Erich Focht}
\IEEEauthorblockA{OpenChip\\
Email: erich.focht@openchip.com}
\and
\IEEEauthorblockN{Marc Casas}
\IEEEauthorblockA{Barcelona Supercomputing Center\\
Universitat Politècnica de Catalunya\\
Email: marc.casas@bsc.es}}

\maketitle

\begin{abstract}
GEneral Matrix Multiplications (GEMMs) are recurrent in high-performance computing and deep learning workloads. 
Typically, high-end CPUs accelerate GEMM workloads with Single-Instruction Multiple Data (SIMD) or vector Instruction Set Architectures (ISAs).
Since these ISAs face significant issues when running GEMM workloads, particularly when dealing with small, tall, or skinny matrices, matrix ISAs have been proposed and implemented by major hardware vendors in the last years.
Although these matrix ISAs deliver larger throughput when running GEMMs than their SIMD/vector counterparts, they are rigid solutions unable to dynamically adapt themselves to application-specific aspects like the data format.
This paper demonstrates that the state-of-the-art matrix ISAs deliver suboptimal performance when running the most commonly used convolution and transformer models.

This paper proposes the {\it Matrix Tile Extension (\mte{})}, the first matrix ISA that 
completely decouples the instruction set architecture from the microarchitecture and seamlessly interacts with existing vector
ISAs.
MTE incurs minimal implementation overhead since it only requires a few additional instructions and a 64-bit Control Status Register (CSR) to keep its state.
Specifically, MTE can i) vectorize GEMMs across the three dimensions $M$, $N$, and $K$; ii) leverage the capacity of the existing vector register file; and iii) decouple the tile shape from the underlying microarchitecture.
MTE achieves speed-ups of \supmtes{} over the best state-of-the-art matrix ISA.
\end{abstract}

\section{Introduction}\label{sec:intro}

GEneral Matrix Multiplications (GEMMs) are ubiquitous in high-performance computing and deep learning workloads~\cite{sze2017efficient}.
Typically, high-end CPUs accelerate GEMM workloads with Single-Instruction Multiple Data (SIMD) or vector Instruction Set Architectures (ISAs)~\cite{georganas2018anatomy,GEMMAVX512,santana2023efficient} to leverage their high-throughput floating-point functional units.
In a push for even higher throughput, major hardware vendors are now incorporating matrix ISAs into CPU architectures with the first implementations released in the last years~\cite{amx, moreira2021matrix, sifivemma, thead}.
These approaches achieve better compute throughput than SIMD/vector ISAs by exploiting specialized Matrix-Multiply-Accumulate (MMA) units.
Ultimately, these matrix extensions complement the existing SIMD/vector ISAs instead of replacing them~\cite{openpower31, amx, armsme, sifivemma}.

The advantages of matrix ISAs do not come for free since the architecture must efficiently handle matrix operands in hardware and expose a concise software/hardware interface.
Solutions to expose matrix operands to hardware include vector register grouping~\cite{openpower31}, tile registers~\cite{amx}, or registers with a capacity of $n^2$ elements, with $n$ being the vector/SIMD register width~\cite{armsme}.
These approaches are tightly coupled with microarchitectures implementing short-vector ISAs~\cite{AVX512,fugaku,moreira2021matrix}.
Although they are effective, these solutions consist in rigid matrix ISAs supporting static matrix tile shapes unable to dynamically adapt themselves to application-specific aspects like the data format. 
This paper demonstrates the state-of-the-art matrix ISAs deliver suboptimal performance when running the commonly used convolution and transformer models.

This paper proposes the {\it Matrix Tile Extension (\mte{})}, the first matrix ISA that completely decouples the instruction set architecture from the microarchitecture.
Additionally, the \mte{} ISA seamlessly interacts with the existing vector ISAs, which makes \mte{} able to leverage vector architecture registers for matrix computations and vector instructions for element-wise operations.
\mte{} incurs minimal implementation overhead since it only requires six additional operations and a 64-bit Control Status Register (CSR) to keep its state. Specifically, \mte{} can i) vectorize GEMMs across the three GEMM loops over $M$, $N$, and $K$; ii) fully exploit the capacity of the vector register file for storing matrices in uniform and mixed-precision scenarios; and iii) decouple the matrix tile geometry from the underlying microarchitecture. By leveraging these three properties, \mte{} achieves significant performance gains with respect to state-of-the-art vector~\cite{AVX512, RVV, SVE} and matrix~\cite{amx, sifivemma} ISAs for a heterogeneous set of GEMM workloads belonging to relevant convolution and transformer models.

Specifically, this paper makes the following contributions:

\noindent
\textbullet~~ It identifies the main performance bottlenecks of state-of-the-art matrix ISAs. These bottlenecks come from poor utilization of the matrix storage space, and also the restricted number of registers these ISAs expose to the compiler.
 
\noindent
\textbullet~~ It proposes \mte{}, the first geometry-agnostic ISA decoupled from the microarchitecture, enabling code portability across implementations without programmer intervention.
The paper also describes two microarchitectures supporting \mte{}, the first one is based on a lean extension of a long vector processor and the second one leverages a systolic array.

\noindent
\textbullet~~ It evaluates the performance of \mte{} considering 75 convolution workloads belonging to deep neural networks~\cite{resnet, inception, vgg, yolo, squeezenet} and 18 transformer workloads obtained from natural language processing models~\cite{vaswani2017attention, brown2020language} and recommendation systems~\cite{sun2019bert4rec, wu2020sse, gupta2020architectural}.
Our evaluation compares \mte{} against two approaches based on a vector ISA featuring 8192- and 16,384-bit vector registers, and two recently proposed matrix ISAs~\cite{amx, sifivemma}.
Our evaluation indicates that \mte{} delivers better performance than these four approaches on top of a high-performance computing architecture.
In particular, \mte{} beats the best state-of-the-art approach, \amx{}~\cite{amx}, by \supmtes{} on average due to software leveraging a larger number of architecturally-visible registers for matrix operands.
\section{Background and Motivation}\label{sec:fundamentals}

This section describes the standard software interface of GEMM operations (Section~\ref{sec:fundamentals:interface}) and presents modern vector architectures (Section~\ref{sec:fundamentals:vector}). Also, this section describes state-of-the-art CPU matrix ISAs (Section~\ref{sec:fundamentals:matrix}), and indicates their shortcomings (Section~\ref{sec:shortcomings}).

\subsection{The Basic Linear Algebra Subprograms (BLAS) Interface}\label{sec:fundamentals:interface}

The BLAS~\cite{BLAS} specification is an ubiquitous standard software interface describing, among others, the GEneral Matrix Multiplication (GEMM).
The GEMM operation is defined as $C \gets \alpha AB + \beta C$ where $A$, $B$, and $C$ are matrices with shapes ($M$,$K$), ($K$,$N$), and ($M$,$N$), respectively, while $\alpha$ and $\beta$ are scalar scaling factors.
Table~\ref{tab:foundations:gemm} describes the BLAS GEMM call interface.
The $M$, $N$, and $K$ parameters define the operand's shape and the geometry of the GEMM computation, which is characterized by the $M \times N \times K$ triplet.
$M$ describes the number of rows in A and C, $N$ specifies the number of columns of $B$ and $C$, and $K$ indicates the number of columns of $A$ and rows of $B$.
The leading dimension offset states the distance between elements in consecutive rows or columns, and decouples the matrix memory organization from the operation shape. The $TRANS\{A,B,C\}$ flags express the memory layout of operands as row-major or col-major.

\begin{table}
    \caption{The BLAS GEMM routine arguments}
    \vspace{-0.25cm}
    \small
    \begin{center}
        \begin{tabular}{ c c c c }
            \toprule
            Parameter & Description \\
            \midrule
            A, B, C & Pointer to the operand matrices \\
            M, N, K & Matrix multiplication size      \\
            LD\{A, B, C\} & Matrices leading dimension offset \\
            TRANS\{A, B, C\} & Matrices memory layout \\
            $\alpha$,$\beta$ & Scaling factors \\
            \bottomrule
        \end{tabular}
        \label{tab:foundations:gemm}
    \end{center}
    \vspace{-0.1cm}
\end{table}

\subsection{SIMD/Vector Support in Modern Processors}\label{sec:fundamentals:vector}

Modern processors expose some form of vector or SIMD programming model as part of an Instruction Set Architecture (ISA) extension to accelerate linear algebra and AI workloads~\cite{AVX512,SVE,RVV}.
To implement such models, processors employ parallel compute units operating on vector registers, delivering several floating-point operations per cycle on data-parallel workloads~\cite{GEMMAVX512,georganas2018anatomy,zhang2018high}.

\subsubsection{Vector ISAs}
Emerging vector ISAs such as the {\it RISC-V Vector Extension (\rvv{})}~\cite{RVV} and the ARM {\it Scalar Vector Extension (SVE)}~\cite{SVE} support Vector Length Agnostic (VLA) programming models~\cite{SVE}, allowing applications to run in vector processors implementing any vector register size without code changes.
The flexibility of VLA programming models has renewed interest in long vector architectures, as evidenced by production systems~\cite{Aurora23} and research on architectures employing vector registers as wide as 16384 bits~\cite{minervini2023vitruvius+,maceiras2022vsa}.
Despite their good performance on HPC workloads~\cite{armejach2018stencil,gomez2021efficiently,Aurora23}, long vector architectures are not able to fully use their floating-point computing capacity on GEMMs coming from computer vision AI tasks~\cite{gupta2023challenges,santana2023efficient}.

\subsubsection{Microarchitecture}
\begin{figure}
\centerline{\includegraphics[width=.7\linewidth]{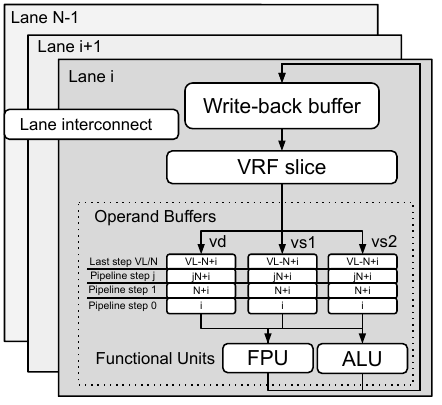}}
\vspace{-0.2cm}
\caption{Structure of the $i$-th lane of a vector unit with N lanes. The vector register file storage is interleaved across the N lanes. Functional units contain one execution pipeline per lane operating on local elements across $VL/N$ steps. The lane interconnect enables communication between lanes.}
\label{img:proposal:hardware-vpu}
\vspace{-0.2cm}
\end{figure}

The defining characteristic of vector arithmetic instructions is that they implement operations that combine the $i$-th element stored in a vector register with the corresponding $i$-th element of another register.
This characteristic drives the design of vector units, which are often implemented with an array of deeply pipelined functional units, or lanes, operating on subvectors~\cite{patterson2020}.
Figure~\ref{img:proposal:hardware-vpu} represents the $i$-th lane of a vector architecture with $N$ lanes that completes a vector instruction over $VL$ elements in $VL / N$ steps.
The vector register file is interleaved across lanes and a lane interconnect enables inter-lane communication to support vector reductions, slides, and similar operations~\cite{Aurora23}.
The vector operand buffers are present on each lane to receive vector element operands from the local Vector Register File (VRF) slice in preparation toward the functional units.
Vector architectures support configurable vector length which disables computation on vector tail elements by scheduling fewer elements into the operand buffers.
In addition, vector architectures support predication, which discards the functional unit outputs for specific elements and thus preserve their values.

\subsection{Matrix ISAs for CPU Architectures}\label{sec:fundamentals:matrix}

General-purpose CPU architectures are gradually incorporating matrix multiplication ISA extensions to accelerate Artificial Intelligence (AI) and High-Performance Computing (HPC) workloads~\cite{amx,sifivemma,openpower31,armsme}.
These systems expose matrix multiplication instructions implemented by a Matrix Multiply-Accumulate (MMA) unit, which leverages the data reuse potential of MMA operations and outperform approaches based purely on SIMD/Vector units.

\subsubsection{The Intel Advanced Matrix Extensions (\amx{})} AMX consists of a dedicated set of eight tile registers with 1KB of storage each, and a tile matrix multiply unit (TMUL) to operate on them~\cite{amx-sdm}.
To use the TMUL, software must first configure the shape of each individual tile register, in terms of the number of active rows and the row size in bytes, by populating a 64-byte CSR using a dedicated instruction.
The \amx{} TMUL unit is a grid of fused multiply-add units able to read and write to tile registers.
\amx{} TMUL instructions support only mixed-precision integer (int8 to int32) and floating-point (bf16 to fp32) operations with a maximum $M \times N \times K$ geometry of $16 \times 16 \times 64$ and $16 \times 16 \times 32$, respectively.

To enforce maximum utilization of the tile storage, TMUL instructions require the $A$ and $C$ matrices to be laid out in row-major layout while the $B$ matrix must undergo a relayout, packing adjacent rows into 32-bit words (two bf16 or four int8 elements).
Element-wise operations like $\alpha$ and $\beta$ scaling required by the BLAS GEMM we describe in Section~\ref{sec:fundamentals:interface} are not supported by \amx{} and must be implemented using either scalar instructions or SIMD approaches like AVX512~\cite{AVX512}.
However, since there is no interface between AVX512 and AMX register files, communication is done through memory before post-processing.

\subsubsection{SiFive Intelligence Custom Extensions (SiFiveInt)}

The most recent custom matrix extension for \rvv{} is SiFiveInt~\cite{sifivemma}, which introduces an MMA instruction involving several 4$\times$4 matrix tiles.
This instruction uses three vector register operands, two bf16 input vectors ($vs1$ and $vs2$), and a single fp32 input/output vector ($vd$), and conceives vector registers as a sequence of independent 4$\times$4 matrix tiles.
The MMA instruction multiplies the first tile in $vs1$, the $A$ operand, to all tiles in $vs2$, the $B$ operand, and accumulates the results into the $vd$ tiles, the $C$ operand.
Despite being a vector-length agnostic extension, the constant 4$\times$4 operand shape for $vs1$ incurs poor hardware utilization on long vector implementations, as the MMA uses just the first 128 bits of the $vs1$ register ($i.e.$, one 4$\times$4 Bfloat16 tile).
Finally, the \sifive{} extension does not define matrix memory move operations and relies on tiled memory layouts for efficient data movement with unit-stride vector load/stores, otherwise requiring expensive vector gather/scatter.

\subsection{Shortcomings of Matrix ISAs}
\label{sec:shortcomings}
To illustrate the shortcomings of state-of-the-art matrix ISAs,
we evaluate the single-core efficiency of the x86 {\it Advanced Matrix Extension (\amx{})}~\cite{amx} and the {\it Advanced Vector Extension (AVX512)}~\cite{AVX512} on an Intel Xeon Platinum 8480+~\cite{xeon8480} processor locked to a frequency of 2.1 GHz.
Our evaluation considers 75 unique convolution layers from computer vision networks~\cite{resnet,vgg,squeezenet,yolo,inception} and 31 GEMM workloads commonly featured on transformer networks~\cite{vaswani2017attention}, GPT~\cite{brown2020language} and BERT-based models for recommendation systems~\cite{sun2019bert4rec, wu2020sse, gupta2020architectural}.

\begin{figure}
\vspace{-0.3cm}
\centerline{\includegraphics[width=.99\linewidth]{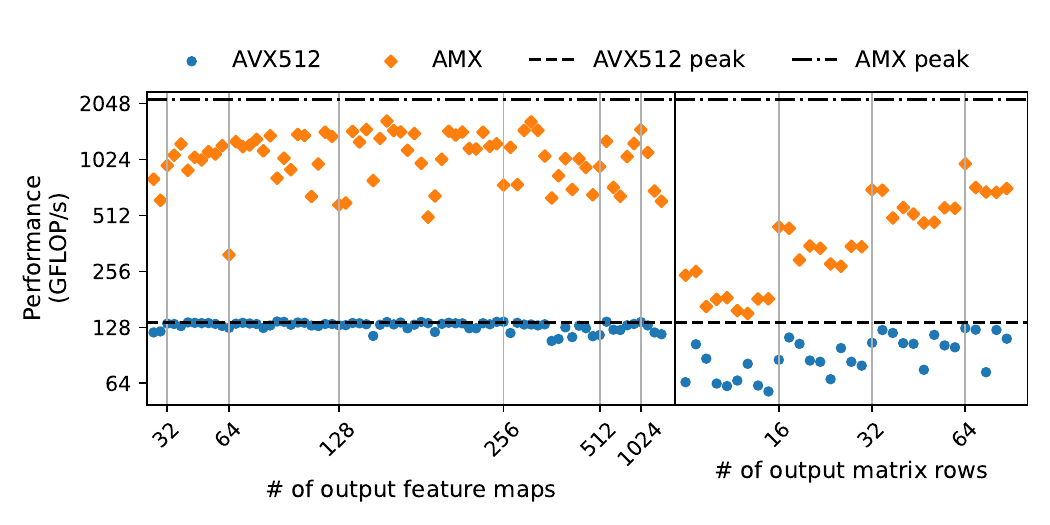}}
\vspace{-0.4cm}
\caption{Single core Intel Xeon Platinum 8480+ performance of vector (AVX512) and matrix (AMX) ISAs during computer vision convolutions and language models GEMMs.}
\label{img:fundamentals:efficiency}
\vspace{-0.2cm}
\end{figure}

Figure~\ref{img:fundamentals:efficiency} shows our results.
The y-axis represents the measured performance in terms of Giga Floating-Point operations per second (GFLOP/s) in logarithmic scale.
The x-axis
on the left subplot
displays all the convolution workloads we consider sorted by the ascending number of output feature maps.
The x-axis on the right subplot showcases the GEMM workloads extracted from transformers and recommendation systems sorted by the ascending number of output matrix rows.
We measure average efficiencies of 35.4\% for AMX and 85.6\% for AVX512 in terms of percentage of peak performance.
While the arithmetic throughput of AMX is 16$\times$ greater than AVX512 (2150 bf16 GOps/s versus 134 fp32 GOps/s), the limited single-core efficiency constrains AMX speedup over AVX512 to a range of 5.7–10$\times$~\cite{amx-accelerate}.

The poor efficiency of \amx{} in terms of floating-point performance with respect to its peak performance is explained by two main factors: i) the fact that \amx{} requires dedicated 1KB registers to store matrix operands reduces the number of architectural registers to only 8, otherwise the matrix register file would be prohibitively large, which limits the potential performance enhancements of compiler-based approaches like loop unrolling; and ii) the rigid $M \times N \times K$ geometry of \amx{}, which is restricted to just $16 \times 16 \times 64$ and $16 \times 16 \times 32$ matrix tile computations and does not match tall and skinny matrix tiles, which usually appear in AI workloads like the ones we represent in Figure~\ref{img:fundamentals:efficiency}. 
In addition, the shapes of GEMM kernels coming from transformer models~\cite{vaswani2017attention, brown2020language} depend on input parameters, which makes it impractical to store input matrices on pre-defined high-performance layouts as it happens on modern direct convolution kernels.
Therefore, we observe comparatively lower AMX performance for GEMMs coming from transformer models as data must be transposed and/or re-ordered via SIMD instructions, stored to memory, and reloaded to AMX tile registers.



This paper proposes a novel matrix ISA extension targeting GEMMs to fix the limitations of current approaches, since i) it uses vector registers to store matrix tiles and therefore does not require dedicated matrix registers to store matrix tiles; ii) it is able to dynamically adapt its tile geometry to workload requirements. 




\section{The Matrix Tile Extension}\label{sec:proposal}

This section describes the Matrix Tile Extension (\mte{}) ISA, which extends the concept of vector length agnosticism~\cite{RVV, SVE} to support a flexible geometry-agnostic matrix instruction set architecture.
\mte{} reuses the vector register file for tile storage, avoiding the need for dedicated registers and matrix-to-vector register moves.

\subsection{Determining Tile Dimensions}\label{sec:proposal:vstate}

This section describes how MTE determines the sizes of $A$, $B$, and $C$ matrix tiles to support a geometry-agnostic matrix instruction set that maximizes the use of vector registers storage capacity. Section~\ref{sec:uniformprecision} describes the scenario where the $A$, $B$, and $C$ matrix operands involved in the GEMM computation are composed of scalar coefficients of the same data type. We call this scenario \textit{Uniform Precision}. Section~\ref{sec:mixedprecision} describes the scenario where matrix $C$ is composed of scalar coefficients represented with a larger data type than matrices $A$ and $B$. We call this scenario \textit{Mixed Precision}, similarly as previous work~\cite{mixedprecision}.

\subsubsection{Uniform Precision}
\label{sec:uniformprecision}

In the uniform precision scenario, all matrices contain scalar coefficients represented with the same data type. In this context, the \rvv{} ISA~\cite{RVV} uses vector registers to store a set of $VLEN/SEW$ elements, where $VLEN$ is the size of the vector registers in bits and $SEW$ indicates the single element bit-width.
The SVE ISA~\cite{SVE} applies a similar approach and expresses the state contained by the $VLEN$ and $SEW$ parameters in terms of predicates. 
For simplicity, and without loss of generality, we express the state required by \mte{} using the \rvv{} nomenclature, although it could be written in terms of any vector-length agnostic ISA.
\mte{} considers 32 vector registers, like \rvv{} and SVE.

\begin{figure}
\centerline{\includegraphics[width=.8\linewidth]{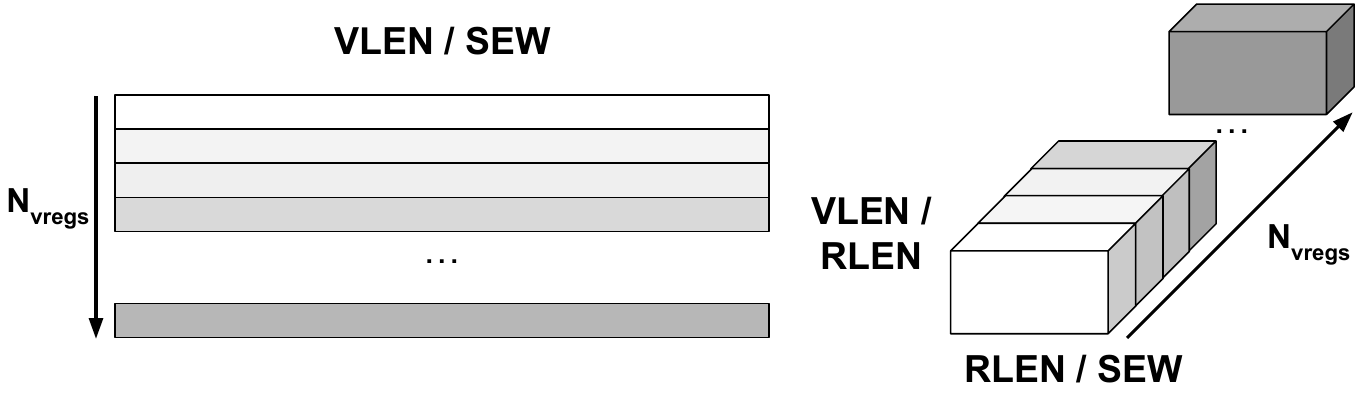}}
\vspace{-0.1cm}
\caption{Rank-1 (left) and rank-2 (right) interpretations of the vector register file. The vector view interprets registers as vectors of VLEN/SEW elements. The rank-2 view interprets registers as VLEN/RLEN $\times$ RLEN/SEW matrices.}
\label{img:proposal:vrfview}
\vspace{-0.1cm}
\end{figure}

\mte{} organizes matrix elements by segmenting the vector registers into $\frac{VLEN}{RLEN}$ matrix rows storing $\frac{RLEN}{SEW}$ elements each, as illustrated in Figure~\ref{img:proposal:vrfview}.
The $RLEN$ parameter is a \mte{} design-time constant informing the tile row size in bits.
Formula~\ref{eq:proposal:shape} expresses the largest hardware matrix geometry considering the $VLEN$, $SEW$, and $RLEN$ parameters.
The product $ROWS \cdot COLS$ is equal to the storage of a vector register in terms of element count, $VLEN/SEW$.

\begin{equation}
\label{eq:proposal:shape}
\begin{split}
ROWS = \frac{VLEN}{RLEN} \textrm{ , }  COLS = \frac{RLEN}{SEW}
\end{split}
\end{equation}

In the uniform precision scenario, \mte{} stores matrix tiles in vector registers in row-major order.
The three equations of Formula~\ref{eq:proposal:dimensions} determine the maximum $M$, $N$, and $K$ dimension sizes supported by a processor with $VLEN$-bits vector registers, and $RLEN$-bits rows for a given value of $SEW$.
The $K$ dimension formula constrains the matrix size to the storage space of a single vector register.
Formula~\ref{eq:proposal:dimensions} guarantees the full use of vector registers storing $C$ tiles regardless of the $VLEN$, $RLEN$, and $SEW$.
Fully utilizing the vector registers storing $A$ and $B$ tiles requires $M = N$ since the $K$ dimension represents the number of rows in $B$ and columns in $A$.


\begin{equation}
\begin{split}
\label{eq:proposal:dimensions}
M = \frac{VLEN}{RLEN} \textrm{ , } N = \frac{RLEN}{SEW} \textrm{ , } K = min(M, N) \\
\end{split}
\end{equation}

\subsubsection{Mixed Precision}
\label{sec:mixedprecision}

In the mixed precision scenario,  matrix $C$ is
composed of scalar coefficients represented with a larger data type
than $A$ and $B$.
This difference implies that two different $SEW$ sizes must be considered in the mixed precision scenario: the $SEW$ of matrices $A$ and $B$, which we call $SEW_i$, and the one of $C$, which we call $SEW_o$.
Having distinct SEW values brings an additional restriction when determining $N$, $i.e.$, $N=min(\frac{RLEN}{SEW_i},\frac{RLEN}{SEW_o}$). Since $SEW_i < SEW_o$, the data size of $C$ determines the value of $N$, and therefore the $B$ tile size is smaller than the vector register capacity.

To overcome this issue, \mte{} uses a transposed layout to store $B$ tiles in vector registers in the mixed precision scenario, $i.e.$, \mte{} organizes and operates on $B$ tiles in a col-major ordering within vector registers.
Using a transposed layout exclusively for $B$ operands alters the semantics of the $K$ dimension which now determines the number of columns of $A$ and $B$ matrices, $i.e.$ $K=\frac{RLEN}{SEW_i}$.
In addition, $N$ now determines both 
the number of rows of $B$ and columns of $C$, therefore $N=min(\frac{VLEN}{RLEN}, \frac{RLEN}{SEW_o})$.
Taking into account that $M=\frac{VLEN}{RLEN}$, we obtain the three equations of Formula~\ref{eq:proposal:dimensions-bt}.
They determine the maximum $M$, $N$, and $K$ dimension sizes involving a transposed $B$ operand and considering the different input and output data sizes, $SEW_i$ and $SEW_o$.

\begin{equation}
\begin{split}
\label{eq:proposal:dimensions-bt}
M = \frac{VLEN}{RLEN} \textrm{ , } N = min(M, \frac{RLEN}{SEW_o}) \textrm{ , } K = \frac{RLEN}{SEW_i} \\
\end{split}
\end{equation}

Therefore, transposing $B$ makes it possible to determine $N$ by considering $SEW_o$ instead of both $SEW_o$ and $SEW_i$.
Because there is no equation in Formula~\ref{eq:proposal:dimensions-bt} involving both $SEW_i$ and $SEW_o$, the inequality $SEW_i < SEW_o$ does not restrict the full use of the vector register length.
For instance, an architecture targeting 32-bit matrix operations with $VLEN$ 8192 and $RLEN$ 512 describes a maximum matrix multiplication geometry of 16x16x16 in a uniform precision scenario with 32-bit data types.
This scenario fully utilizes the vector register capacity (256 elements) on all operands, according to Formula~\ref{eq:proposal:dimensions}.
The same architecture executing mixed-precision operations with $SEW_o = 32$ and $SEW_i = 16$ describes a maximum geometry of 16x16x32, also using full vector registers capacity (256 output elements, 512 input elements), according to Formula~\ref{eq:proposal:dimensions-bt}.

\begin{table}
    \caption{64-bit CSR for the Matrix Tile Extension}
    \vspace{-0.3cm}
    \begin{center}
        \begin{tabular}{ l l r }
            \toprule
            Name  & Description & Bits             \\
            \midrule
            t[m,n,k]    & Tile dimension shapes & 36\\
            ttype[i,o]  & Input/output matrix tile types     & 8\\
            rlenb       & RLEN in Bytes         & 12 \\
            Reserved    & additional data       & 8 \\
            \bottomrule
        \end{tabular}
        \label{tab:proposal:csr}
    \end{center}
\end{table}

\subsection{Control Status Registers}\label{sec:proposal:csr}

\mte{} stores all the state defining the tile's geometry in a 64-bit Control Status Register (CSR) described in Table~\ref{tab:proposal:csr}.
The $tm$, $tn$, and $tk$ fields store the current hardware geometry settings for the $M$, $N$, and $K$ dimensions with a maximum dimension size of $2^{12} = 4096$ elements.
The $ttypei$ and $ttypeo$ are two 4-bit fields specifying traits of matrix input and output operands.
The $ttype$ fields use 2 bits to specify the $SEW$ of each operand tile, which can be either 8, 16, 32, or 64 bits.
The remaining 2 bits on the $ttype$ fields codify how to handle elements on inactive columns and rows.
One option is to leave the inactive bits untouched (undisturbed policy in the RISC-V nomenclature~\cite{RVV}).
Alternatively, the software is responsible for not accessing such elements as they can be dirty (agnostic policy in RISC-V).
The $rlenb$ CSR holds the $RLEN$ size in bytes.
The remaining bits are reserved for \mte{} extensions.

\subsection{\mte{} Instructions}\label{sec:instructions}

This section describes the 19 \mte{} instructions organized in five instruction groups, as seen in Table~\ref{tab:proposal:instructions}.

\begin{table}
    \caption{Matrix Tile Extension instructions}
    \vspace{-0.3cm}
    \begin{center}
        \resizebox{\columnwidth}{!}{%
        \begin{tabular}{ l l l }
            \toprule
            Instruction     & Arguments    & Description \\
            \midrule
            tss[m,n,k]      & rd, rs1, ttypeio      & tile set shape [M,N,K] \\
            t\{t\}l[a,b,c,bt]  & vd, rs1, rs2 & [A,B,C,B$^T$] tile \{transposed\} load \\
            t\{t\}sc  & vd, rs1, rs2 & C tile \{transposed\} store \\
            t\{f\}\{w\}mul       & vd, vs1, vs2 & tile \{FP\} \{widening\} dot product \\
            tvmask[a,b,c,bt]   & vd, rs1      & set vector mask \\
            \bottomrule
        \end{tabular}
        }
        \label{tab:proposal:instructions}
    \end{center}
\end{table}

\subsubsection{Configuring the tile geometry}\label{sec:model:cfg}

The $tssm$, $tssn$, and $tssk$ instructions configure the $M$, $N$, and $K$ GEMM matrix dimensions by updating the respective $tm$, $tn$, or $tk$ CSR fields.
The instructions take the requested dimension size from the application in $rs1$ and return the updated CSR value, the granted dimension size, in $rd$.
The return value is the minimum between the application request and the maximum dimension size allowed by the underlying microarchitecture, discussed in Sections~\ref{sec:uniformprecision} and~\ref{sec:mixedprecision}, 
and the data-width of input and output matrices, $SEW_i$, and $SEW_o$.
The $ttypeio$ is a 3-bit immediate encoding of the input and output matrix element widths, $SEW_i$ and $SEW_o$, for configuring the $ttype$ CSR fields.

\subsubsection{Matrix data movement}\label{sec:model:memory}

The tile load/store, $tl$/$ts$, and their transposed variants $ttl$/$tts$, move GEMM matrix operands between memory and the vector register file.
In the general case, these instructions access blocks of up to $RLEN$ consecutive bits found at constant strides in memory.
The load instructions are encoded with the GEMM operand type they produce, $A$, $B$, $C$, or $B^T$ to specify the relevant CSR dimension shape fields to the microarchitecture for instruction decoding.
For instance, $C$ tile operations require the CSR $tm$, and $tn$ fields to compute the $C$ tile shape.
The special $t\{t\}lbt$ instruction indicates that the $B$ operand is organized in a col-major layout within the vector register state.
Finally, \mte{} defines store instructions exclusively for the $C$ matrix operand.

\mte{} memory instructions require two scalar parameters: The base memory address from $rs1$, and the leading dimension offset from $rs2$, expressed in bytes.
The \mte{} operands on memory instructions are directly associated to the BLAS software interface parameters pairs ($A$, $lda$), ($B$, $ldb$), and ($C$, $ldc$).
The special case with stride zero corresponds to row or column broadcast operations where a single row/column value in memory is replicated across all vector register rows/columns.
\mte{} implementations may optimize row/column broadcasts when $rs2$ describes the 0-stride scenario.

\subsubsection{Matrix Multiplication}\label{sec:model:multiply}

The $tfmul$ and $tmul$ matrix tile multiply instructions compute the product of the $A$ and $B$ matrix tiles and accumulate the result to a $C$ tile matrix, $i.e.$, they are MMA instructions.
The former instruction operates on floating-point data types and the latter on integers.
These instructions require three vector register operands, the $C$ tile in $vd$, the $A$ tile in $vs1$, and the $B$ tile in $vs2$.
The $tm$, $tn$, and $tk$ \mte{} CSR fields determine the operation shape.
The $tfwmul$ and $twmul$ are the mixed-precision variants of $tfmul$ and $tmul$, respectively, and interpret the $B$ operand in terms of a col-major layout, as Section~\ref{sec:mixedprecision} indicates.

\begin{figure}
\centerline{\includegraphics[width=.7\linewidth]{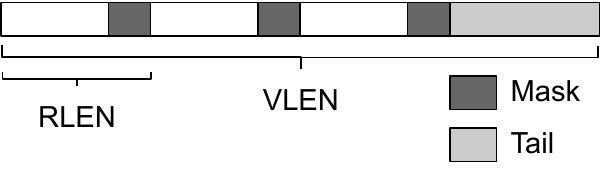}}
\vspace{-0.2cm}
\caption{Vector processing of matrices in vector registers. White and dark strips represent active and inactive elements respectively. The vector length deactivates inactive matrix rows at the vector tail, and the vector mask deactivates inactive columns at each row spanning $RLEN$ bits.}
\label{img:proposal:vmode}
\end{figure}

\subsubsection{Vector processing mode}\label{sec:model:vector}

\mte{} employs vector instructions to implement element-wise operations on matrix data stored in vector registers.
This interaction removes the need for data movement between memory or register files.
For instance, the vector ISA can implement matrix additions and scalar-matrix multiplications, like the ones of BLAS GEMM, with vector additions ($vfadd.vv$) and vector-scalar multiplications ($vfadd.vf$), respectively.

Figure~\ref{img:proposal:vmode} illustrates how to disable computations on specific matrix elements stored inside the vector register state using the programmable vector length and vector mask features of vector architectures.
Setting up the vector instructions to operate on matrix tiles requires: i) setting the vector length to guarantee that all active rows are covered; and ii) creating a vector mask to disable computations on inactive columns at each row.
In the context of \rvv{}, the $vsetvl$ instruction~\cite{RVV} performs the first step, gathering the application vector length and type from source scalar registers.
The application can read the architecture row size in the $rlenb$ CSR field to compute the vector length for $vsetvl$ in a geometry-agnostic way.
Finally, for the vector mask, \mte{} introduces the $tvmask$ instructions to create a vector mask for either $A$, $B$, $C$, or $B^T$ tiles based on the active tile geometry settings.

\begin{algorithm}[t]
    \caption{\mte{} SGEMM $C \gets \alpha AB + \beta C$}\label{alg:model:mgemm}
    \textbf{Input:} $A, B, C, M, N, K, LDA, LDB, LDC, \alpha, \beta$ \\
    \textbf{Output:} $C$
    \begin{algorithmic}[1]
        \State {$sm, sn, sk = 0$}
        \State {$erow = read\_csr(rlenb) / sizeof(float)$}
            \For {$m=0,M, m += sm$}
                \State $sm = tssm(M-m)$
                \State {$vl = sm * erow$}
                \For {$n=0,N, n += sn$}
                    \State {$sn = tssn(N-n)$}
                    \State {$gvl = vsetvl(vl, e32)$}
                    \State {$vm = tvmaskc(gvl)$}
                    \State $c = vbroadcast(0.0f, gvl)$
                    \For {$k=0,K, k += sk$}
                        \State $sk = tssk(K-k)$
                        \State $a = tla(\&A[m*LDA+k], LDA, tm, tk)$
                        \State $b = tlb(\&B[k*LDB+n], LDB, tk, tn)$
                        \State $c = tfmul(c, a, b, tm, tn)$
                    \EndFor
                    \State $t = tlc(\&C[m*LDC+n], LDC, tm, tn)$
                    \State $c = vfmul\_vf\_mask(c, \alpha, vm, gvl)$
                    \State $c = vfmacc\_vf\_mask(t, \beta, vm, gvl)$
                    \State $tsc(c, \&C[m*LDC+n], LDC, tm, tn)$
                \EndFor
            \EndFor
    \end{algorithmic}
\end{algorithm}

\subsection{BLAS GEMM Algorithm using \mte{}}
\label{sec:SGEMM}

Algorithm~\ref{alg:model:mgemm} shows a BLAS SGEMM $C \gets \alpha AB + \beta C$ routine implemented with \mte{} assuming the \rvv{} ISA extension is supported.
Line 2 computes the number of elements per matrix row in hardware.
Lines 4, 7, and 12 configure the GEMM geometry at each iteration.
The granted dimension sizes are stored in the $sm$, $sn$, and $sk$ variables for incrementing the $M$, $N$, and $K$ loop iterators in Lines 3, 6, and 11.
The Lines 4-5, and 8-9 configure the hardware to operate on the $C$ matrix.
Line 10 initializes the $C$ output matrix accumulator with zeros using the \rvv{} scalar-vector broadcast instruction.
Lines 13-15 depict the computational kernel of the algorithm, consisting of two tile load instructions, from $A$ and $B$, and a tile multiply operation.
Lines 16-18 use standard masked vector arithmetic operations to implement the $\alpha$ and $\beta$ scalar-matrix multiplications on the $C$ matrix loaded from memory.
Typically, this algorithm is optimized by unrolling the $M$ and/or $N$ loops to reuse the $B$ and/or $A$ matrix tiles loaded into registers in operations across multiple independent $C$ output tiles within the $K$ loop.
This optimization serves the purpose of hiding the latencies of the tile load operations by increasing the compute density of the innermost loop and exposing more independent work to the hardware.
The $K$ loop may also be unrolled to avoid frequent CSR writes.

\subsection{Summary of the MTE ISA}\label{sec:architecture:sota}

\mte{} proposes to reuse the vector register file to store matrix tile operands.
This strategy makes it possible to eliminate the need for extra architectural state to support MMA instructions.
Similarly to \amx{}, \mte{} also uses a CSR for controlling the tile shapes but simplifies the configuration process by enabling only three simultaneous matrix shapes for $A$, $B$, and $C$, which are programmed by software using a few scalar instructions.
\mte{} also provides a set of instructions to configure the vector programming model based on the \mte{} state, allowing a seamless transition from matrix to vector processing mode without the need for memory or register moves.
In addition, \mte{} defines a geometry-agnostic programming model independent of the matrix multiplication shapes supported by the microarchitecture, while defining a standard way for the software to expose the required geometry to the hardware.

\section{Architecture Support for MTE Instructions}\label{sec:architecture}

This section describes the architecture support required by MTE.
Section~\ref{sec:architecture:tmul} describes two possible microarchitecture implementations of the MTE MMA instructions. 
Section~\ref{sec:architecture:memory} describes the support required by the MTE memory instructions, while Section~\ref{sec:architecture:other} discuses additional aspects to fully support MTE.

\begin{figure}
\centerline{\includegraphics[width=\linewidth]{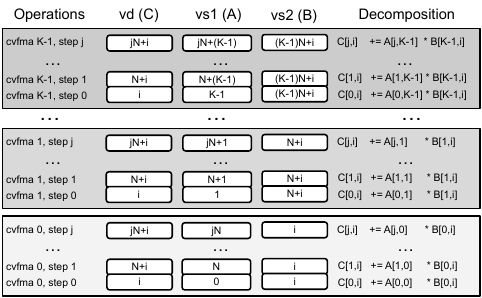}}
\caption{Operand buffers contents within one vector lane during a matrix multiplication decomposed into $K$ $cvfma$ operations. The left and right sections depict the correlation between $cvfma$ steps and matrix semantics. The middle section shows the operand buffer contents.}
\label{img:proposal:hardware-vpu-buffers}
\end{figure}

\subsection{Supporting the $tfmul$ and $tmul$ Instructions}\label{sec:architecture:tmul}

When computing $tfmul$ and $tmul$ MMA operations, described in Section~\ref{sec:model:multiply}, inputs from the $A$ and $B$ tiles contribute to multiple $C$ output elements.
In this section, we describe the architecture support required by MTE in the context of i) a dedicated systolic array; and ii) a standard vector processor.

\subsubsection{Systolic array:}
A common approach to implement MMA operations is through dedicated systolic arrays consisting of a grid of functional units, such as those found in Intel AMX or Google TPUs~\cite{amx,choquette2023nvidia,jouppi2021ten,maceiras2022vsa}. These architectures directly read from tile registers, delivering high throughput per operation. However, a notable drawback is their specialization for MMA operations, which can result in unused resources and limited datatype support. For example, AMX and TPU hardware only support up to 16-bit input operands. Extending support to 32-bit or 64-bit inputs would incur significant area overheads, as a multiplier area scales quadratically with mantissa bit width~\cite{areaFMA,area8bit}.

\subsubsection{Vector processor:}
To exploit data reuse at the register level in a vector architecture, we propose breaking down MMA instructions into a sequence of new vector micro-instructions called Component Vector Fused-Multiply-Accumulate ($cvfma$). The $cvfma$ instructions behave like standard vector FMAs except for the lane control logic, which modifies the operand buffer filling to implement the data reuse patterns required by matrix multiplication. Without loss of generality, this section focuses on supporting $tmul$; the same method applies to $tfmul$.

The $tmul$ instruction is decoded into $K$ $cvfma$ operations, based on the \mte{} $tk$ CSR field. Each $cvfma$ operation processes a vector length of $M \cdot RLEN$ bits, determined by the \mte{} $tm$ CSR field, which disables computation on tail elements corresponding to inactive tile rows. To avoid computations on inactive columns across the $N$ dimension, the architecture employs an implicit vector predication mask on $cvfma$ instructions, derived from the \mte{} $tn$ CSR field. This technique is compatible with software predication masks, combining them with a logical $and$ operation to disable computations on arbitrary elements.

Figure~\ref{img:proposal:hardware-vpu-buffers} illustrates the $tmul$ decomposition and shows the operand buffer contents for the $i$-th vector lane across all $K$ $cvfma$ instructions. The $vd$ operand buffer, containing the $C$ tile, is filled similarly to regular vector architectures, as indicated in Figure~\ref{img:proposal:hardware-vpu}. The $vs2$ operand buffer, mapped to the $B$ tile, holds a single lane-local element for all $cvfma$ steps. This pattern, called implicit broadcast, is already used in scalar-vector arithmetic instructions like the \rvv{} $vfmacc.vf$ instruction~\cite{RVV}. No inter-lane communication is required for the $vs2$ operand.
Finally, the $A$ tile operand requires an access pattern unsupported by standard vector instructions. Specifically, each $cvfma$ operation in the sequence requires accessing the operand buffer of a specific lane: the first $cvfma$ accesses lane zero, the second lane one, and so forth. To support this access pattern, \mte{} can leverage the interconnect between the different vector lanes to move $vs1$ operands between lanes.

Our approach supports the $tfmul$ and $tmul$ MMA instructions using a state-of-the-art vector architecture as a basis.
Each lane manages all computations within a matrix column, the pipeline steps make it possible to iterate over the matrix rows, the operand buffers support the vertical flow of $B$ values, and the lane interconnect implements the flow of $A$ values. Reusing the existing vector engine hardware for computation improves hardware utilization while supporting wide datatypes (32 and 64 bit) seamlessly, necessary in classic HPC applications~\cite{linpack, hpcg}.

\subsection{Memory Instructions}
\label{sec:architecture:memory}

The general use case of \mte{} matrix memory instructions can be broken down into a series of unit-stride vector load/stores of up to $RLEN$ bits in a constant stride.
If the stride value matches the architecture $RLEN$, all the matrix tile data is contiguous in memory and therefore the hardware can map the matrix memory operations to a simple unit-stride vector load/store.
A $tl$/$ttl$ instruction with zero stride, possibly signaled with the zero register ($x0$ in the case \rvv{}), triggers a row/column broadcast, replicating a sequence of $RLEN$ bits to all $VLEN$ bits if the vector register.

\subsection{Other Support}
\label{sec:architecture:other}

The $tss$ instructions for configuring the hardware GEMM shape are similar to $vsetvl$ instructions in \rvv{} that, in turn, configure the vector instruction traits using CSRs.
Similar to some existing vector HPC architectures, we support register renaming on the \mte{} CSRs to enable speculation and allowing matrix instructions to be executed out-of-order.
\section{Evaluation Methodology}\label{sec:experiments}

This section describes our evaluation methodology in terms of system architecture, workloads, and considered vector/matrix ISAs.

\subsection{System Architecture}\label{sec:experiments:hardware}

Tables~\ref{tab:experiments:architecture},~\ref{tab:experiments:architecture:vector}, and~\ref{tab:experiments:architecture:systolic} depict the system, vector processing unit, and the systolic array accelerator architectures we consider in our experimentation campaign.
The system and systolic array parameters match an Intel Xeon Platinum 8480+ processor~\cite{xeon8480} core, the latest x86 server-class processors implementing the \amx{} extension.
The Vector Processing Unit (VPU) parameters resemble a single NEC SX-Aurora 20B vector accelerator core~\cite{yamada2018vector} augmented with one vector unit for a total of four.
Each unit has 2048-bit vector lanes, $i.e.$ delivers 64 32-bit FMA instructions per cycle, resulting in a peak performance of 512 Single-Precision (SP) FLOP/cycle, $i.e.$, 1024 GFLOP/s at 2.0GHz.

\begin{table}
\caption{System Configuration}
    \vspace{-0.4cm}
    \begin{center}
    \resizebox{\columnwidth}{!}{%
        \begin{tabular}{ l l }
            \toprule
            Scalar core               & Out of order, 2 GHz, 512 ROB entries \\
            Issue/Commit width & 6 \\
            L1D & 48KB, 8-way, 4cc, 10-entry MSHR, LRU\\
            L1I & 32KB, 8-way, 4cc, 10-entry MSHR, LRU\\
            L2 & 2MB, 8-way, 26cc, 256-entry MSHR, LRU, 128-byte lines\\
            Main memory bandwidth   & 191.25 GB/s per core \\
            Main memory latency     & 110ns \\
            \bottomrule
        \end{tabular}
        }\label{tab:experiments:architecture}
    \end{center}
    \vspace{-0.25cm}
\end{table}

\begin{table}
    \caption{Vector processing unit}
    \vspace{-0.4cm}
    \begin{center}
    \resizebox{\columnwidth}{!}{%
        \begin{tabular}{ l l }
            \toprule
            Vector registers   & 32 (40 physical) \\
            Vector length      & 8192 bits for all approaches but {\it \slv{}} (16384 bits). \\
            Vector lane width  & 2048 bits \\
            Vector units       & 4 \\
            Throughput         & 512 FLOP (SP) per cycle\\
            \bottomrule
        \end{tabular}
    }\end{center}\label{tab:experiments:architecture:vector}
    \vspace{-0.25cm}
\end{table}

\begin{table}
    \caption{Systolic array accelerator}
    \vspace{-0.4cm}
    \begin{center}
    \resizebox{\columnwidth}{!}{%
        \begin{tabular}{ l l }
            \toprule
            Tile registers & 8 (24 physical) on \samx{}, 32 (40 physical) on \smtes{} \\
            Tile length    & 8192 bits \\
            Row length     & 512 bits \\
            Throughput     & 512 FLOP (SP) per cycle\\
            
            \bottomrule
        \end{tabular}
    }\end{center}\label{tab:experiments:architecture:systolic}
    \vspace{-0.3cm}
\end{table}

In this evaluation, we consider systolic and vector systems delivering the same peak throughput, that is, 512 FLOP/cycle and 1024 GFLOP/s.
This methodology makes it possible to compare ISAs with different hardware GEMM implementations while highlighting the impacts of the ISA when expressing GEMMs.

\subsection{Workloads}
\label{sec:workloads}
This section highlights the workloads used in our evaluation as well as our code generation method.

\subsubsection{Code generation}\label{sec:experiments:codegen}

We extend oneDNN~\cite{onednn}, an architecture-agnostic kernel library providing AI operators under a standard software interface, with convolution and GEMM algorithms for all the evaluated architectures described in Section~\ref{sec:techniques}.
Popular interfaces like Pytorch~\cite{paszke2019pytorch} and Tensorflow~\cite{tensorflow2016} employ oneDNN as a CPU backend for compute-intensive operations such as convolutions and GEMMs.
oneDNN CPU primitives for x86 and Arm architectures use the Xbyak Just-in-Time (JIT) assembler~\cite{xbyak,xbyakArm} to produce micro-kernels specialized for the operation parameters and target architecture on the fly.
We develop \textit{rvjit}, a RISC-V JIT assembler library similar to Xbyak to dynamically generate micro-kernels for \rvv{} and our proposed extensions.

We implement direct convolution kernels for vector and matrix architectures following state-of-the-art recipes for SIMD architectures~\cite{santana2023efficient, georganas2018anatomy, ferrari2023advancing}.
The direct algorithm employs a tiled matrix memory layout for both activation and weight tensors, and reduces the convolution to a series of matrix tile multiplications.
The tiled memory layouts enable unit-stride vector/memory operations.
Since input and output activation tensors have a consistent organization, no data reshapes are required as the layout propagates across network layers.

Our JIT-generated micro-kernels for vector and \mte{} ISAs use system balance equations~\cite{georganas2018anatomy,santana2023efficient} to tune performance aspects like loop unrolling and tiling to the architecture VLEN, RLEN, and cache size parameters.
The micro-kernels vectorize the $N$ loop, unroll the $M$ and $N$ loop, and apply cache-blocking on the $K$ loops of the GEMM formulation.
When applied to convolutions, we map the minibatch, output feature map, and input feature map dimensions to the $M$, $N$, and $K$ GEMM matrix dimensions.

\begin{table*}[t]
    \caption{Evaluated architectures}
    \vspace{-0.25cm}
    \begin{center}
    \resizebox{\textwidth}{!}{%
        \begin{tabular}{ l | c c | c c | c c c c c }
            \toprule
            \textbf{Architecture} &
            \multicolumn{2}{|c}{\textbf{Constants}} &
            \multicolumn{2}{|c|}{\textbf{Register file size}} &
            \multicolumn{5}{c}{\textbf{tfmul (matrix) or vfma (vector) instruction support}}
            \\

            Acronym &
            VLEN &
            RLEN &
            Architecture &
            Physical &
            Static latency &
            Dynamic latency &
            Geometry &
            VPUs &
            Systolic array
            \\

            \midrule
            
            \sv{} &
            8192 &
            - &
            32 &
            40 &
            20 &
            4 &
            1$\times$256$\times$1 &
            4 &
            -
            \\
            
            \slv{} &
            16384 &
            - &
            32 &
            40 &
            20 &
            8 &
            1$\times$512$\times$1 &
            4 &
            -
            \\
            
            \ssf{} &
            8192 &
            2048 &
            32 &
            40 &
            28 &
            16 &
            16$\times$16$\times$4 &
            4 &
            No
            \\
            
            \samx{} &
            8192 &
            512 &
            8 &
            24 &
            36 &
            16 &
            16$\times$16$\times$16 &
            2 &
            Yes
            \\

            \smtes{} &
            8192 &
            512 &
            32 &
            40 &
            36 &
            16 &
            16$\times$16$\times$16 &
            2 &
            Yes
            \\
            
            \smtev{} &
            8192 &
            512 &
            32 &
            40 &
            36 &
            64 &
            16$\times$16$\times$16 &
            4 &
            No
            \\
            \bottomrule
        \end{tabular}
    }
    \end{center}\label{tab:experiments:methods}
    \vspace{-0.2cm}
\end{table*}

\subsubsection{Convolution Workloads}\label{sec:experiments:convolutions}

We evaluate convolution workloads from the popular computer vision networks ResNet~\cite{resnet}, Inception~\cite{inception}, VGG~\cite{vgg}, yolo~\cite{yolo}, and SqueezeNet~\cite{squeezenet}.
We obtain the models from torchvision~\cite{marcel2010torchvision}, an open-source machine vision package, execute them using Pytorch~\cite{paszke2019pytorch}, and collect the convolution parameters by inspecting the library calls to oneDNN.
We reproduce the Pytorch convolution calls for each layer using benchdnn~\cite{benchdnn}.
Our evaluation covers \nexperiments{} unique convolution operations including pointwise, spatial, strided, and padded convolutions with square and non-square kernel geometry, and utilizes 32-bit floating-point datatypes.
Finally, we define a minibatch size of 16 for our experiments, creating sufficiently large problems to saturate all the evaluated architectures.

\subsubsection{Transformer Workloads}\label{sec:experiments:transformers}

We evaluate \mte{} considering 18 GEMM workloads commonly featured on transformer networks~\cite{vaswani2017attention}, GPT~\cite{brown2020language} and BERT-based models for recommendation systems~\cite{sun2019bert4rec, wu2020sse, gupta2020architectural}.
These GEMMs cover the linear projections necessary for multi-head attention, the scalar-dot-product function used by the attention block, and the feed-forward network~\cite{vaswani2017attention}.
We evaluate these operations on an inference scenario with small query sizes (16, and 32), two different encoding dimensions, $d_{model}$, of 512 and 768 elements with 8 and 12 heads, respectively, and feed-forward operations with 2048 hidden connections.

\subsection{Evaluated architectures}
\label{sec:techniques}

This section describes the ISAs we consider in our evaluation. All ISAs run on top of the systems specified in Section~\ref{sec:experiments:hardware} except \amx{}, which runs on an Intel Xeon Platinum 8480+~\cite{xeon8480} processor.

{\it \sv{}:} uses the \rvv{} ISA to support a direct convolution approach~\cite{georganas2018anatomy, santana2023efficient, ferrari2023advancing} vectorizing the $N$ loop, and unrolling the $M$ loop of the GEMM kernel.
Our JIT assembler generates \rvv{} code and uses all 32 architecture registers for unrolling.
The maximum vector length is 8192 bits or 256 32-bit elements.

{\it \slv{}:} it employs the same approach as \sv{} except for the maximum vector length, which is set to 16384 bits.
In this scenario, vector instructions operate on up to 512 32-bit elements.

{\it \ssf{}:} it uses a matrix kernel with a custom MMA instruction similar to SiFiveIntelligence~\cite{sifivemma}, adapted to fp32 operands, which multiplies independent 4$\times$4 matrix tiles spanning 512 consecutive bits within a vector register.
We consider a long vector architecture with $VLEN$ 8192 and use \mte{} instructions to emulate \sifive{} semantics by defining $RLEN$ as 2048, resulting in a capacity for 16 4$\times$4 matrix tiles and a maximum GEMM geometry of 4$\times$64$\times$4.
We organize the tensor data in 4$\times$4 tiles to enable stride-1 memory requests which corresponds to the best use case for this ISA.

{\it \smtev{}:} uses the proposal described in Section~\ref{sec:proposal} and implements the GEMM instructions as Section~\ref{sec:model:vector} indicates on top of the VPU described in Table~\ref{tab:experiments:architecture:vector}.
We consider an \mte{} implementetion with a $VLEN$ of 8192 and $RLEN$ of 512, resulting in a GEMM geometry of 16$\times$16$\times$16, similar to \amx{}.
The JIT generator uses all 32 \rvv{} registers for loop unrolling.

{\it \smtes{}:} It employs the same approach as \smtev{} but implements GEMM instructions using the systolic array described in Table~\ref{tab:experiments:architecture:systolic}.
Both \smtes{} and \smtev{} have the same peak performance since \smtev{} can use all vector units to compute four simultaneous GEMMs whereas \smtes{} computes one tile at a time. 

{\it \samx{}:} It reproduces the semantics of the x86 AMX ISA~\cite{amx} using \mte{}.
This strategy uses at most eight architectural registers for code generation.

{\it \amx{}:} We consider the x86 {\it Advanced Matrix Extension (\amx{})} ISA~\cite{amx} implemented on the Intel Xeon Platinum 8480+~\cite{xeon8480} processor locked to a frequency of 2.1 GHz.
Our evaluation leverages Intel's optimized kernels in oneDNN v3.5~\cite{onednn}, integrated as the CPU backend within PyTorch v2.2~\cite{paszke2019pytorch}. 

Table~\ref{tab:experiments:methods} describes the evaluated architectures architectural parameters, register file sizes, and support for matrix or vector instructions.
We do not include \amx{} since its architecture parameters are very similar as the ones of \samx{}.
The latency fields, expressed in cycles, corresponds to the static and dynamic instruction cost, described in Section~\ref{sec:experiments:software}.
All considered strategies display the same peak performance of 1024 GFLOPs/s considering 32-bit datatypes.


\begin{table}
    \caption{Physical register file area of each architecture}
    \vspace{-0.2cm}
    \begin{center}
    \addtolength{\tabcolsep}{-0.4em}
    \resizebox{\columnwidth}{!}{%
        \begin{tabular}{ccccccc}
            \toprule
             & \sv{} & \slv{} & \ssf{} & \samx{} & \smtes{} & \smtev{} \\
            \midrule
            $mm^2$& $1.66$ & $4.15$ & $1.66$ & $1.65$ & $1.66$ & $1.66$ \\
            \bottomrule
        \end{tabular}
    }
    \end{center}
\label{tab:experiments:areaPRF}
    \vspace{-0.25cm}
\end{table}

\subsection{Area Cost}
\label{sec:areaandenergy}
This section describes the area cost of the architectures we discuss in Section~\ref{sec:techniques}. Section~\ref{sec:energy} describes the energy consumption of these architectures.
We estimate the energy and area consumption for different hardware components using McPAT 1.3~\cite{mcpat09} and PCACTI~\cite{pcacti, FINCACTI, pcacti-5nm}, respectively. We incorporate the enhancements proposed by Xi et al.~\cite{Xi.2015.HPCA}. These enhancements improve accuracy by modeling additional core structures and correcting erroneous modeling assumptions. Our analysis assumes a 5nm FinFET technology node to estimate the area cost~\cite{pcacti-5nm}.

Table~\ref{tab:experiments:areaPRF} details the physical register file area for each architecture we describe in Section~\ref{sec:techniques}.
The \slv{} architecture has the largest register file area, as it includes 40 2KB registers, whereas the other approaches use 1KB registers, as shown in Table~\ref{tab:experiments:methods}. Architectures with 40 1KB registers require less than half the area of \slv{}. Among all the considered approaches, \samx{} has the smallest register file area, as it contains only 24 1KB registers.
Using the same methodology,  we determine that 
the primary contributor to the area cost is the register file for all considered architectures. 



\subsection{Simulation methodology and validation}\label{sec:experiments:software}

We evaluate the architectures from Section~\ref{sec:experiments:hardware} using a trace-driven micro-architecture simulator that models physical register allocation, memory movement across cache levels, and dynamic instruction latency based on active vector length.
Vector and matrix instructions are simulated with two cost components: i) a static, non-blocking, front-end latency paid after decode and before reserving compute resources which can be overlapped with the execution of other instructions, and ii) a dynamic latency tied to vector length and compute throughput that blocks the compute resource.

\begin{figure}
\centerline{\includegraphics[width=\linewidth]{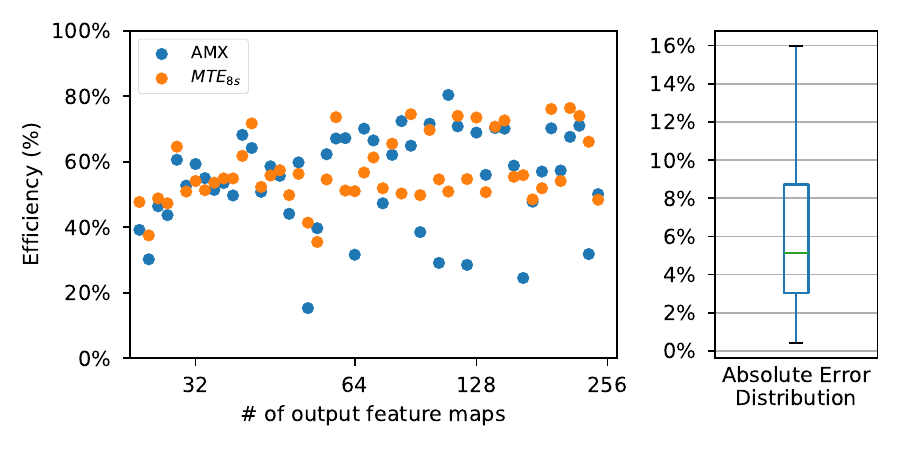}}
\vspace{-0.3cm}
\caption{Measured and simulated performance of convolution dataset on an Intel Xeon Platinum 8480+ (left) and the absolute error distribution (right).}
\label{img:experiments:amx-simulator}
\vspace{-0.25cm}
\end{figure}

We validate our simulation infrastructure by comparing its results for the \samx{} scenario, which replicates the x86 AMX ISA (Section~\ref{sec:techniques}), against real AMX executions on an Intel Xeon Platinum 8480+ processor~\cite{xeon8480}.
Figure~\ref{img:experiments:amx-simulator} presents data for 52 unique convolutions with $OC \leq 256$ (Section~\ref{sec:experiments:convolutions}).
The left subplot shows simulated vs. measured efficiency, and the right shows absolute error.
The simulator matches \amx{} peak performance and showcases a median error of 5.0\% and first/third quartiles at 3.0\% and 8.7\%.
We observe that \samx{} delivers slightly better performance than \amx{} on problems with larger $OC$ values.
We attribute this behavior to cache conflict misses taking place on \amx{} oneDNN algorithm on the 8480+ processor's cascading cache architecture~\cite{amx-measure}.

\section{Evaluation}\label{sec:results}

We evaluate the seven approaches described in Section~\ref{sec:techniques} when running the workloads of Sections~\ref{sec:experiments:convolutions} and~\ref{sec:experiments:transformers}.
All approaches run on top of the system described in Table~\ref{tab:experiments:architecture}, except \amx{}, which runs on an Intel Xeon Platinum~\cite{xeon8480} processor.
Section~\ref{sec:results:efficiency} evaluates the performance of these approaches in terms of the peak performance percentage they achieve.
Section~\ref{sec:energy} evaluates our proposals in terms of energy consumption.
Section~\ref{sec:results:instructions} presents an analysis on the retired instruction count.

\begin{figure*}[ht]
\vspace{-0.4cm}
\centerline{\includegraphics[width=1\linewidth]{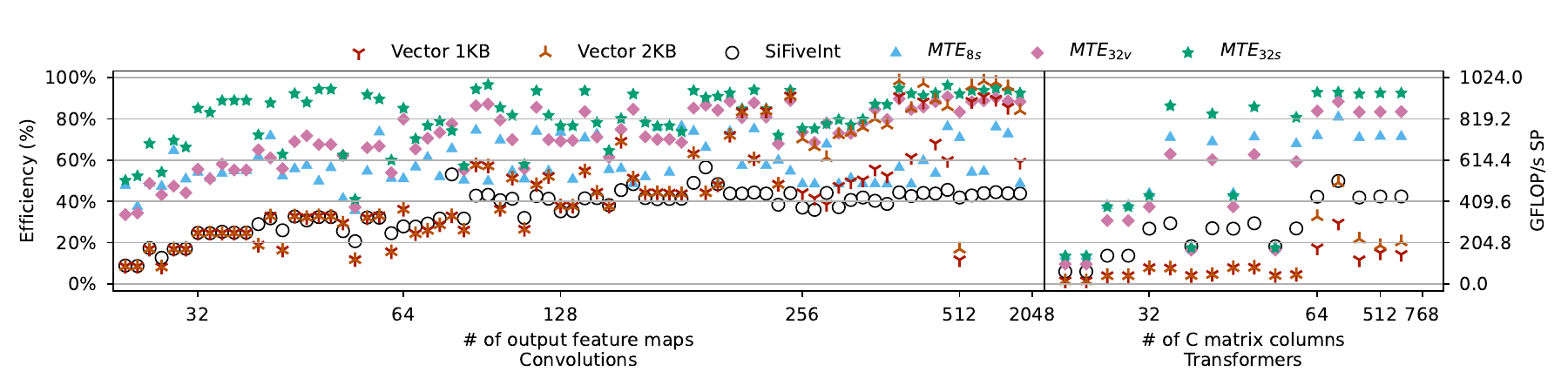}}
\vspace{-0.45cm}
\caption{Percentage of the peak performance (efficiency) obtained by convolution and GEMM kernels. Convolution and GEMM workloads are displayed in ascending order concerning the number of output feature maps and $C$ matrix columns respectively.}
\vspace{-0.0cm}
\label{img:experiments:efficiency}
\end{figure*}

\subsection{Performance}
\label{sec:results:efficiency}

Figure~\ref{img:experiments:efficiency} shows the peak performance percentage achieved by each approach across 75 convolutions and 18 transformer workloads, organized by increasing output feature maps ($OC$) or output matrix columns ($N$), which are most suited for vectorization.
We classify Workloads into six categories based on $OC$/$N$ sizes: i) 1–32, ii) 33–64, iii) 65–128, iv) 129–256, v) 257–512, and vi) 513–2048.

\sv{} and \slv{} show increasing average efficiencies across categories I-IV (\xevi{} to \xeviv{}) because their performance is limited by the data-parallelism available in the vectorized dimension, $i.e.$, $OC$, or $N$.
For convolution workloads in categories V and VI, \slv{} reaches \celvv{}, and \celvvi{} efficiency, respectively, by exploiting its wider vector length.
In contrast, some GEMM shapes do not equally divide the \slv{} maximum vector length on transformer workloads ($e.g.$ $N$ = 768, VL = 512), causing hardware under-utilization on the last $N$ loop iterations and lower efficiencies of \telvv{} and \telvvi{} for transformer GEMMs of categories V-VI.

The \ssf{} kernel obtains average efficiencies of \xesfi{}, \xesfii{}, \xesfiii{}, \xesfiv{}, \xesfv{}, and \xesfvi{} on categories I-VI considering both convolution and transformer workloads.
Despite exploiting the parallelism of the entire GEMM loop nest, \ssf{} fails to deliver floating-point throughput close to the peak.
The \ssf{} ISA defines a hardware GEMM shape of 4$\times$64$\times$4 when implemented on processors with a $VLEN$ of 8192 bits.
This geometry results in a relatively small 4$\times$4 $A$ matrix operand, causing a register state under-utilization and comparatively lower arithmetic density.

\samx{}, which reproduces the semantics of the x86 \amx{} ISA, successfully vectorizes all GEMM loops and, therefore, it exposes larger data-level parallelism than \sv{}, \slv{}, and \ssf{} for categories I-IV. 
For categories V and VI, the \slv{} approach delivers larger performance than \samx{} for convolution workloads since the larger $OC$ size enables \slv{} to fully use the wider vector length, as well as maximizing loop unrolling across the 32 available vector registers, while \amx{} is limited to eight.
 
\smtev{} and \smtes{} successfully use the full vector length in all workloads, leading to better performance than \sv{} and \slv{}, especially when dimensions $OC$ and $N$ are less than 256.
\smtev{} obtains \xevmtei{} average efficiency on category I and between \xevmteii{} and \xevmtevi{} in categories II-VI.
\smtes{} obtains \xesmtei{} average efficiency on category I and between \xesmteii{} and \xesmtevi{} in categories II-VI.
\smtev{} and \smtes{} are \supmtev{} and \supmtes{} faster, on average, than \samx{} since the former approaches are able to leverage a larger number of architectural registers than the latter.
The \vlonger{} approach reaches \mte{} performance when the vectorized dimension contains: i) 512 elements or more elements; and ii) a number of elements that is multiple of 512, as it is the case for category VI convolutions.
\smtev{} obtains geometric mean speedups of \suv{}, \sulv{}, \susf{} and \susamx{} over \vlong{}, \vlonger{}, \sifive{}, and \samx{} while \smtes{} achieves \ssuv{}, \ssulv{}, \ssusf{} and \ssusamx{}.



\subsubsection{End-to-end evaluation}
\label{sec:endtoend}

This section evaluates MTE in the context of end-to-end inference on complete AI models.
We conduct our analysis using PyTorch 2.5.1~\cite{paszke2019pytorch} and consider computer vision models from the torchvision package v0.20.1~\cite{torchvision2016}, including: SqueezeNet~\cite{squeezenet}, Inception~\cite{inception}, and ResNet50~\cite{resnet}.
Additionally, we evaluate the language models BERT~\cite{vaswani2017attention} (configured for the fill-mask task) and GPT-2~\cite{brown2020language} (configured for text generation) obtained via the transformers package v4.48.3~\cite{wolf-etal-2020-transformers}.


\begin{figure}
\vspace{-0.3cm}
\centerline{\includegraphics[width=\linewidth]{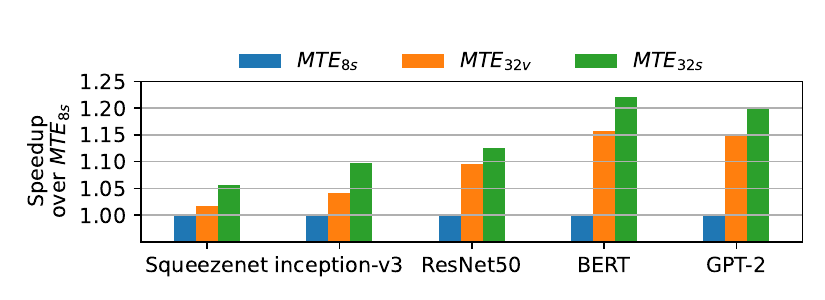}}
\vspace{-0.35cm}
\caption{Application-level speedup of \smtev{} and \smtes{}, considering computer vision and language models.}
\label{img:experiments:end-to-end}
\vspace{-0.4cm}
\end{figure}



Figure~\ref{img:experiments:end-to-end} shows the performance speedup of \smtes{} and \smtev{} over \samx{}, which models the AMX ISA~\cite{amx} semantics.
\sv{}, \slv{}, and \ssf{} are omitted due to inferior performance.
\smtes{} achieves speedups of 1.05$\times$, 1.09$\times$, and 1.13$\times$ on SqueezeNet, Inception, and ResNet50, respectively, while \smtev{} sees 1.02$\times$, 1.04$\times$, and 1.10$\times$ gains.
For BERT and GPT-2, \smtes{} reaches 1.20$\times$ and 1.22$\times$, and \smtev{} gets 1.15$\times$ and 1.16$\times$. These gains stem from MTE’s acceleration of convolution workloads in computer vision models and GEMMs in language models.
BERT and GPT-2 show the highest gains, as they spend 76.16\% and 67.04\% of inference time on GEMMs, compared to 37.22\%, 51.36\%, and 48.92\% for SqueezeNet, Inception, and ResNet50 on convolutions.

\subsubsection{Comparison between MTE and AMX}
\label{sec:MTEvsAMX}

\begin{figure}
\vspace{-0.1cm}
\centerline{\includegraphics[width=1\linewidth]{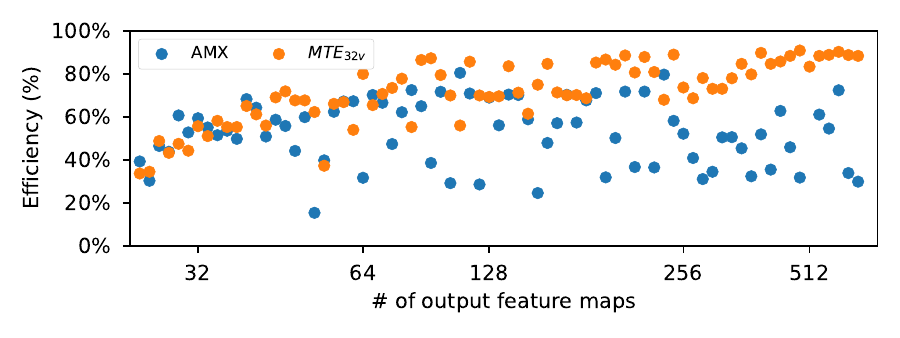}}
\vspace{-0.4cm}
\caption{Convolution efficiency obtained by \amx{} compared to MTE$_{32v}$ simulated performance data.}
\vspace{-0.1cm}
\label{img:experiments:amx-mte}
\end{figure}

Figure~\ref{img:experiments:amx-mte} shows the \amx{} performance for the considered convolution workloads on the 8084+ processor.
The figure also shows the performance of the \smtev{} approach on the architecture described by Tables~\ref{tab:experiments:architecture} and~\ref{tab:experiments:architecture:vector}.
\amx{} and \smtev{} obtain average efficiencies of 52.8\% and 68.1\%, respectively, defining a \smtev{} speedup of 1.29$\times$ over \amx{}.
\smtev{} obtains better performance than \amx{} primarily when $OC \ge 64$, since \smtev{} leverages its 32 architecture registers for unrolling in this scenario.

Specifically, MTE enables greater degrees of unrolling for the loop over the $M$ dimension in Algorithm~\ref{alg:model:mgemm}, which exposes more independent $tfmul$ instructions (line 15) and improves reuse of the $b$ tile (line 14).
To perform such software optimizations, the algorithm needs: i) a large $M$ dimension size, or $OC$ for convolutions; and ii) a larger number of architectural registers to hold multiple $C$ and $A$ operands.
The latter is not fulfilled by \amx{}, which results in a lower performance than \mte{} on operations with sufficiently large $M$ and $OC$ dimension sizes.


\begin{figure}
\centerline{\includegraphics[width=1\linewidth]{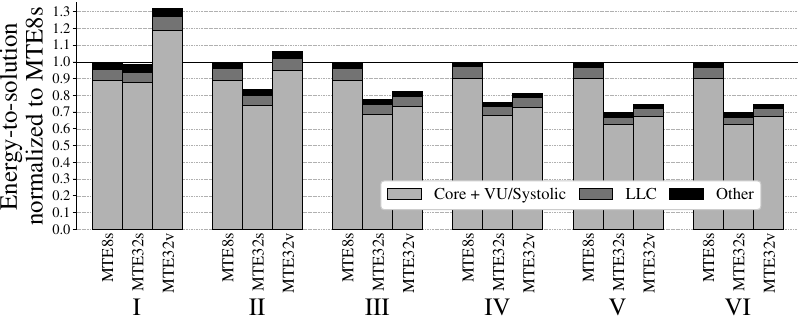}}
\vspace{-0.2cm}
\caption{
Energy-to-solution per category for \smtev{} and \smtes{}, normalized to \samx{}. Components: \texttt{core + VU/Systolic} includes the scalar core, all functional units, and register files; \texttt{LLC} is the last-level cache; and \texttt{other} primarily consists of network-on-chip and memory controllers.}
\vspace{-0.2cm}
\label{img:energy-to-solution}
\end{figure}

\begin{table}
    \caption{Retired vector/matrix instructions reduction for each workload considering the \sv{} baseline.}
    \begin{center}
    \vspace{-0.3cm}
        \resizebox{\columnwidth}{!}{%
        \begin{tabular}{ l r r r r r }
            \toprule
            Category ($OC$ or $N$) & \slv{} & \ssf{} & \samx{} & \smtev{} & \smtes{} \\
            \midrule
            I)   1-32     & 1.00 & 5.97	& 36.40 & 37.22 & 37.22 \\
            II)  33-64    & 1.00 & 5.87	& 17.48 & 18.55 & 18.55 \\
            III) 65-128   & 1.00 & 3.69	& 8.95  & 11.37 & 11.37 \\
            IV)  129-256  & 1.00 & 2.78	& 5.57  & 7.89  & 7.89  \\
            V)   257-512  & 2.00 & 2.76	& 4.95  & 7.88  & 7.88  \\
            VI)  513-2048 & 1.81 & 2.44	& 4.67  & 6.92  & 6.92  \\
			Arithmetic average & 1.24 & 4.05 & 12.38	& 14.31 & 14.31 \\
            \bottomrule
        \end{tabular}
        }
        \label{tab:results:instr_reduction}
    \end{center}
    \vspace{-0.2cm}
\end{table}

\subsection{Energy Consumption}
\label{sec:energy}
This section evaluates the energy consumption of the \smtev{}, \smtes{}, and \samx{} architectures presented in Section~\ref{sec:techniques} following the methodology we describe in Section~\ref{sec:areaandenergy} and considering the workload categories we define in Section~\ref{sec:results:efficiency}.
We do not present results for \sv{}, \slv{}, and \ssf{} since their performance is worse than \smtev{}, \smtes{}, and \samx{}.
Figure~\ref{img:energy-to-solution} shows the geometric mean of energy-to-solution for different hardware components, normalized to \samx{}. In all evaluated configurations, energy consumption is primarily driven by the vector register file, accounting for 77.01\%, 76.59\%, and 77.06\% of the total in \samx{}, \smtes{}, and \smtev{}, respectively. \smtes{} and \smtev{} outperform \samx{} in terms of energy-to-solution by 30.0\% and 25.4\%, respectively, for  workloads belonging to Category VI.

\subsection{Vector/matrix dynamic instruction count}
\label{sec:results:instructions}

Table~\ref{tab:results:instr_reduction} shows the reduction in the retired vector/matrix instruction count considering the \vlong{} baseline on the evaluated convolution and transformer workload groups.
\vlong{} and \vlonger{} trigger the same number of vector instructions when the vectorized dimensions, $OC$ and $N$, have fewer than 256 element.
Otherwise, the \vlonger{} approach requires up to 2$\times$ less instructions than \vlong{}.
\sifive{}, \samx{}, \smtev{}, and \smtes{} require 4.05$\times$, 12.38$\times$, 14.31$\times$, and 14.31$\times$ fewer instructions than \vlong{} on average, respectively.
The largest differences consist of workloads with $OC$ (or $N$) $\leq 256$, where \vlong{} and \vlonger{} are unable to fully use the vector length.
Even for scenarios featuring the largest $OC$ values, \smtev{} and \smtes{} require 1.34$\times$ less instructions than \samx{} as they leverage a larger architecture register file for loop unrolling and run the workloads with fewer micro-kernel calls.
\section{Related Work}\label{sec:related}

The {\it Open Power ISA v3.1 (OpenPower)}~\cite{openpower31} introduces outer product instructions for mixed-, single-, double-precision, and integer matrices, using 512-bit accumulators and register-to-register transfers between accumulators and 128-bit vector registers.
The new outer product instructions sources $A$ and $B$ operands from vector registers and $C$ operands from the larger accumulator registers.
To minimize overheads, the IBM Power10 processor~\cite{moreira2021matrix} repurposes half of the register file for accumulators, avoiding added architectural state and context switch costs.
In contrast, \mte{} eliminates the need for register transfers and matrix accumulators, while offering a geometry-agnostic programming model.

The {\it XuanTie Matrix Multiply Extension (MME)} proposes a detached matrix programming model for RISC-V with eight new dedicated matrix registers~\cite{thead}.
MME implementations selecte the $RLEN$ value from a set of valid row bit-lengths settings to determine both the maximum number of rows, $RLEN/32$, and columns, $RLEN/SEW$.
MME defines matrix shape configuration instructions, 2D memory moves, MMA instructions, but also vector-matrix register-register move, as well as many element-wise instructions on matrix register ($e.g$., data type conversions).
Compared to MME, \mte{} reuses the vector registers, and leverages existing \rvv{} instructions for element-wise operations to reduce the implementation overhead.

The {\it Arm Scalable Matrix Extension (SME)}~\cite{armsme} defines a new matrix storage state of size $VLEN \times VLEN$, where $VLEN$ is the vector register length of the Scalable Vector Extension (SVE) ISA.
SME defines an outer product instruction that regards the matrix storage as a number of independent tile registers ($e.g.$ 4 tiles for fp32) employed as accumulators during the outer product of two SVE vectors.
Besides the outer product, SME includes 2d memory moves, mixed vector to multi-vector operations, and others.
\mte{} differs from SME by i) using existing architectural state for accumulators; and ii) implementing MMAs as matrix dot products, which matches with long vector architectures and other contexts where the $VLEN^2$ architectural matrix register is prohibitively expensive.

The {\it Hopper architecture}~\cite{choquette2023nvidia} is the fourth generation of NVIDIA GPUs augmented with tensor cores to accelerate deep learning workloads via 4$\times$4$\times$8 mixed-precision, and 4$\times$4$\times$4 single-precision, matrix multiplication instructions.
Recent work showcases the benefits of explicit tensor core programming via the Warp Matrix Multiply Accumulate (WMMA) API, which exposes tile shape configuration, matrix load/stores, and multiplication, to accelerate cross-correlation~\cite{fujita2023calculation} and convolutions~\cite{liu2021optimizing}.
Developing efficient kernels for tensor cores relies on adapting the application to the predetermined geometry in the micro-architecture and managing the GPU memory subsystem, which incentives the use of kernel libraries such as cuDNN~\cite{jorda2019performance} maintained by the vendors.

Table~\ref{tab:relatedwork} summarizes the main aspects of the matrix ISAs outlined in this section and Section~\ref{sec:architecture:sota}.
The \mte{} approach is the first matrix ISA that is agnostic to both the tile shape and tile size to decouple the ISA from the underlying microarchitecture.
In addition, \mte{} does not require dedicated registers to store matrix tiles as it relies on the vector register file, and efficiently vectorizes GEMMs across the three dimensions $M$, $N$, and $K$.
\mte{} enables the seamless interplay of vector and matrix instructions, which supports software kernel fusion between GEMM-based workloads and common post-operation like batchnorm and activations.

\begin{table}
\caption{Matrix Extensions Summary}
\vspace{-0.3cm}
\begin{center}
\resizebox{0.49\textwidth}{!}{%
\begin{tabular}{@{}lcrrrr@{}}
\toprule
\textbf{ISA} & \textbf{Tile Size} & \textbf{FP32 Tile} & \textbf{2D Memory} & \textbf{Dedicated}\\
\textbf{Name} & \textbf{(\#Bits)} & \textbf{Shape (\#Elements)} & \textbf{Instructions} & \textbf{Registers}\\
\midrule
OpenPower~\cite{openpower31} & 512 & 4$\times$4 & No & Yes \\
MME~\cite{thead} & 512-16384 & Agnostic & Yes & Yes \\
SME~\cite{armsme} & $VLEN^2$ & Agnostic (Square) & Yes & Yes\\
Tensor Cores~\cite{choquette2023nvidia} & 512 & 4$\times$4 & Yes & -\\
\amx{}~\cite{amx} & 8192 & 16$\times$16 & Yes & Yes\\
\sifive{}~\cite{sifivemma} & 512-16384 & 4$\times$($VLEN/128$) & No & No \\
\mte{} (Section~\ref{sec:proposal}) & Agnostic & Agnostic & Yes & No \\
\bottomrule
\end{tabular}
}
\label{tab:relatedwork}
\end{center}
\vspace{-0.3cm}
\end{table}
\section{Conclusion}\label{sec:conclusion}

This paper demonstrates the limitations of existing vector and matrix ISAs when dealing with GEMM-based workloads.
The paper links the shortcomings of state-of-the-art approaches when dealing with tall, skinny, or small matrices to the under-utilization of the vector register file, and proposes a lean matrix tile extension, \mte{}, to tackle this issue.
\mte{} reinterprets vector registers as matrix tile operands that are manipulated by a novel geometry-agnostic matrix ISA. 
\mte{} decouples the matrix instruction set architecture from the microarchitecture and leverages existing vector instructions for element-wise operations.

We evaluate \mte{} as well as state-of-the-art matrix and vector ISAs considering 75 convolution and 18 transformer workloads extracted from modern deep learning workloads. 
\mte{} obtains average speedups of \ssuv{}, \ssulv{}, \ssusf{} and \ssusamx{} over vector ISA with 8192- and 16384-bit vector registers, and the state-of-the-art \sifive{} and \amx{} ISAs, respectively.


\section*{Acknowledgment}

This work has received funding from ‘Future of Computing, a Barcelona Supercomputing Center and IBM initiative’ (2023) and has been partially supported by the Spanish Ministry of Science and Innovation MCINAEI/10.13039/501100011033 (contract PID2019-107255GB-C21) and ESF Investing in your future, the Generalitat of Catalunya (contract 2021-SGR00763).
Adrià Armejach is a Serra Hunter Fellow and has been partially supported by the Grant IJCI-2017-33945 funded by MCIN/AEI/10.13039/501100011033.
The authors thank the Departament de Recerca i Universitats de la Generalitat de Catalunya for supporting the Research Group "Performance understanding, analysis, and simulation/emulation of novel architectures" (Code: 2021 SGR 00865).

\bibliographystyle{IEEEtran}

\begin{thebibliography}{10}
\providecommand{\url}[1]{#1}
\csname url@samestyle\endcsname
\providecommand{\newblock}{\relax}
\providecommand{\bibinfo}[2]{#2}
\providecommand{\BIBentrySTDinterwordspacing}{\spaceskip=0pt\relax}
\providecommand{\BIBentryALTinterwordstretchfactor}{4}
\providecommand{\BIBentryALTinterwordspacing}{\spaceskip=\fontdimen2\font plus
\BIBentryALTinterwordstretchfactor\fontdimen3\font minus
  \fontdimen4\font\relax}
\providecommand{\BIBforeignlanguage}[2]{{%
\expandafter\ifx\csname l@#1\endcsname\relax
\typeout{** WARNING: IEEEtran.bst: No hyphenation pattern has been}%
\typeout{** loaded for the language `#1'. Using the pattern for}%
\typeout{** the default language instead.}%
\else
\language=\csname l@#1\endcsname
\fi
#2}}
\providecommand{\BIBdecl}{\relax}
\BIBdecl

\bibitem{sze2017efficient}
V.~Sze, Y.-H. Chen, T.-J. Yang, and J.~S. Emer, ``Efficient processing of deep
  neural networks: A tutorial and survey,'' \emph{Proceedings of the IEEE},
  vol. 105, no.~12, pp. 2295--2329, 2017.

\bibitem{georganas2018anatomy}
E.~Georganas, S.~Avancha, K.~Banerjee, D.~Kalamkar, G.~Henry, H.~Pabst, and
  A.~Heinecke, ``Anatomy of high-performance deep learning convolutions on simd
  architectures,'' in \emph{SC18: International Conference for High Performance
  Computing, Networking, Storage and Analysis}.\hskip 1em plus 0.5em minus
  0.4em\relax IEEE, 2018, pp. 830--841.

\bibitem{GEMMAVX512}
\BIBentryALTinterwordspacing
R.~Lim, Y.~Lee, R.~Kim, and J.~Choi, ``An implementation of matrix---matrix
  multiplication on the intel knl processor with avx-512,'' \emph{Cluster
  Computing}, vol.~21, no.~4, p. 1785–1795, dec 2018. [Online]. Available:
  \url{https://doi.org/10.1007/s10586-018-2810-y}
\BIBentrySTDinterwordspacing

\bibitem{santana2023efficient}
A.~d.~L. Santana, A.~Armejach, and M.~Casas, ``Efficient direct convolution
  using long simd instructions,'' in \emph{Proceedings of the 28th ACM SIGPLAN
  Annual Symposium on Principles and Practice of Parallel Programming}, 2023,
  pp. 342--353.

\bibitem{amx}
Intel, \emph{Intel Architecture Optimization Reference Manual}, 2023,
  \url{https://www.intel.com/content/www/us/en/content-details/671488/intel-64-and-ia-32-architectures-optimization-reference-manual-volume-1.html}.

\bibitem{moreira2021matrix}
J.~E. Moreira, K.~Barton, S.~Battle, P.~Bergner, R.~Bertran, P.~Bhat,
  P.~Caldeira, D.~Edelsohn, G.~Fossum, B.~Frey \emph{et~al.}, ``A matrix math
  facility for power isa (tm) processors,'' \emph{arXiv preprint
  arXiv:2104.03142}, 2021.

\bibitem{sifivemma}
SiFive, ``Sifive intelligence extensions documentation,'' 2024,
  \url{https://www.sifive.com/documentation}.

\bibitem{thead}
T.-H. Semiconductor, ``T-head risc-v matrix extension specification,'' 2024,
  \url{https://github.com/T-head-Semi/riscv-matrix-extension-spec}.

\bibitem{openpower31}
O.~P. Foundation, ``The power instruction set architecture v3.1,'' 2024,
  \url{https://openpowerfoundation.org/specifications/isa/}.

\bibitem{armsme}
ARM, ``The scalable matrix extension (sme), for armv9-a,'' 2024,
  \url{https://developer.arm.com/documentation/ddi0616}.

\bibitem{AVX512}
A.~Sodani, R.~Gramunt, J.~Corbal, H.~Kim, K.~Vinod, S.~Chinthamani, S.~Hutsell,
  R.~Agarwal, and Y.~Liu, ``Knights landing: Second-generation intel xeon phi
  product,'' \emph{IEEE Micro}, vol.~36, no.~02, pp. 34--46, mar 2016.

\bibitem{fugaku}
M.~Sato, Y.~Ishikawa, H.~Tomita, Y.~Kodama, T.~Odajima, M.~Tsuji, H.~Yashiro,
  M.~Aoki, N.~Shida, I.~Miyoshi, K.~Hirai, A.~Furuya, A.~Asato, K.~Morita, and
  T.~Shimizu, ``{Co-Design for A64FX Manycore Processor and "Fugaku"},'' ser.
  SC '20.\hskip 1em plus 0.5em minus 0.4em\relax IEEE Press, 2020.

\bibitem{RVV}
T.~R.-V. Foundation, ``The risc-v vector extension,'' 2024,
  \url{https://github.com/riscv/riscv-v-spec/releases/download/v1.0/riscv-vspec-1.0.pdf}.

\bibitem{SVE}
N.~Stephens, S.~Biles, M.~Boettcher, J.~Eapen, M.~Eyole, G.~Gabrielli,
  M.~Horsnell, G.~Magklis, A.~Martinez, N.~Premillieu, A.~Reid, A.~Rico, and
  P.~Walker, ``The arm scalable vector extension,'' \emph{IEEE Micro}, vol.~37,
  no.~02, pp. 26--39, mar 2017.

\bibitem{resnet}
K.~He, X.~Zhang, S.~Ren, and J.~Sun, ``Deep residual learning for image
  recognition,'' in \emph{Proceedings of the IEEE Conference on Computer Vision
  and Pattern Recognition}, 2016, pp. 770--778.

\bibitem{inception}
C.~Szegedy, V.~Vanhoucke, S.~Ioffe, J.~Shlens, and Z.~Wojna, ``Rethinking the
  inception architecture for computer vision,'' in \emph{Proceedings of the
  IEEE conference on computer vision and pattern recognition}, 2016, pp.
  2818--2826.

\bibitem{vgg}
K.~Simonyan and A.~Zisserman, ``Very deep convolutional networks for
  large-scale image recognition,'' \emph{arXiv preprint arXiv:1409.1556}, 2014.

\bibitem{yolo}
J.~Redmon, S.~Divvala, R.~Girshick, and A.~Farhadi, ``You only look once:
  Unified, real-time object detection,'' in \emph{Proceedings of the IEEE
  conference on computer vision and pattern recognition}, 2016, pp. 779--788.

\bibitem{squeezenet}
F.~N. Iandola, S.~Han, M.~W. Moskewicz, K.~Ashraf, W.~J. Dally, and K.~Keutzer,
  ``Squeezenet: Alexnet-level accuracy with 50x fewer parameters and< 0.5 mb
  model size,'' \emph{arXiv preprint arXiv:1602.07360}, 2016.

\bibitem{vaswani2017attention}
A.~Vaswani, N.~Shazeer, N.~Parmar, J.~Uszkoreit, L.~Jones, A.~N. Gomez,
  {\L}.~Kaiser, and I.~Polosukhin, ``Attention is all you need,''
  \emph{Advances in neural information processing systems}, vol.~30, 2017.

\bibitem{brown2020language}
T.~Brown, B.~Mann, N.~Ryder, M.~Subbiah, J.~D. Kaplan, P.~Dhariwal,
  A.~Neelakantan, P.~Shyam, G.~Sastry, A.~Askell \emph{et~al.}, ``Language
  models are few-shot learners,'' \emph{Advances in neural information
  processing systems}, vol.~33, pp. 1877--1901, 2020.

\bibitem{sun2019bert4rec}
F.~Sun, J.~Liu, J.~Wu, C.~Pei, X.~Lin, W.~Ou, and P.~Jiang, ``Bert4rec:
  Sequential recommendation with bidirectional encoder representations from
  transformer,'' in \emph{Proceedings of the 28th ACM international conference
  on information and knowledge management}, 2019, pp. 1441--1450.

\bibitem{wu2020sse}
L.~Wu, S.~Li, C.-J. Hsieh, and J.~Sharpnack, ``Sse-pt: Sequential
  recommendation via personalized transformer,'' in \emph{Proceedings of the
  14th ACM conference on recommender systems}, 2020, pp. 328--337.

\bibitem{gupta2020architectural}
U.~Gupta, C.-J. Wu, X.~Wang, M.~Naumov, B.~Reagen, D.~Brooks, B.~Cottel,
  K.~Hazelwood, M.~Hempstead, B.~Jia \emph{et~al.}, ``The architectural
  implications of facebook's dnn-based personalized recommendation,'' in
  \emph{2020 IEEE International Symposium on High Performance Computer
  Architecture (HPCA)}.\hskip 1em plus 0.5em minus 0.4em\relax IEEE, 2020, pp.
  488--501.

\bibitem{BLAS}
L.~S. Blackford, A.~Petitet, R.~Pozo, K.~Remington, R.~C. Whaley, J.~Demmel,
  J.~Dongarra, I.~Duff, S.~Hammarling, and G.~Henry, ``An updated set of basic
  linear algebra subprograms (blas),'' \emph{ACM Transactions on Mathematical
  Software}, vol.~28, no.~2, pp. 135--151, 2002.

\bibitem{zhang2018high}
J.~Zhang, F.~Franchetti, and T.~M. Low, ``High performance zero-memory overhead
  direct convolutions,'' in \emph{International Conference on Machine
  Learning}.\hskip 1em plus 0.5em minus 0.4em\relax PMLR, 2018, pp. 5776--5785.

\bibitem{Aurora23}
\BIBentryALTinterwordspacing
K.~Takahashi, S.~Fujimoto, S.~Nagase, Y.~Isobe, Y.~Shimomura, R.~Egawa, and
  H.~Takizawa, ``Performance evaluation of a next-generation sx-aurora tsubasa
  vector supercomputer,'' in \emph{High Performance Computing: 38th
  International Conference, ISC High Performance 2023, Hamburg, Germany, May
  21–25, 2023, Proceedings}.\hskip 1em plus 0.5em minus 0.4em\relax Berlin,
  Heidelberg: Springer-Verlag, 2023, p. 359–378. [Online]. Available:
  \url{https://doi.org/10.1007/978-3-031-32041-5\_19}
\BIBentrySTDinterwordspacing

\bibitem{minervini2023vitruvius+}
F.~Minervini, O.~Palomar, O.~Unsal, E.~Reggiani, J.~Quiroga, J.~Marimon,
  C.~Rojas, R.~Figueras, A.~Ruiz, A.~Gonzalez \emph{et~al.}, ``Vitruvius+: an
  area-efficient risc-v decoupled vector coprocessor for high performance
  computing applications,'' \emph{ACM Transactions on Architecture and Code
  Optimization}, vol.~20, no.~2, pp. 1--25, 2023.

\bibitem{maceiras2022vsa}
M.~V. Maceiras, M.~W. Azhar, and P.~Trancoso, ``Vsa: A hybrid vector-systolic
  architecture,'' in \emph{2022 IEEE 40th International Conference on Computer
  Design (ICCD)}.\hskip 1em plus 0.5em minus 0.4em\relax IEEE, 2022, pp.
  368--376.

\bibitem{armejach2018stencil}
A.~Armejach, H.~Caminal, J.~M. Cebrian, R.~Gonz{\'a}lez-Alberquilla,
  C.~Adeniyi-Jones, M.~Valero, M.~Casas, and M.~Moret{\'o}, ``Stencil codes on
  a vector length agnostic architecture,'' in \emph{Proceedings of the 27th
  International Conference on Parallel Architectures and Compilation
  Techniques}, 2018, pp. 1--12.

\bibitem{gomez2021efficiently}
C.~G{\'o}mez, F.~Mantovani, E.~Focht, and M.~Casas, ``Efficiently running spmv
  on long vector architectures,'' in \emph{Proceedings of the 26th ACM SIGPLAN
  Symposium on Principles and Practice of Parallel Programming}, 2021, pp.
  292--303.

\bibitem{gupta2023challenges}
S.~R. Gupta, N.~Papadopoulou, and M.~Peric{\`a}s, ``Challenges and
  opportunities in the co-design of convolutions and risc-v vector
  processors,'' in \emph{Proceedings of the SC'23 Workshops of The
  International Conference on High Performance Computing, Network, Storage, and
  Analysis}, 2023, pp. 1550--1556.

\bibitem{patterson2020}
D.~A. Patterson and J.~L. Hennessy, \emph{Computer organization and design
  RISC-V edition: the hardware software interface}.\hskip 1em plus 0.5em minus
  0.4em\relax Morgan kaufmann, 2020.

\bibitem{amx-sdm}
Intel, \emph{Intel Software Development Manual}, 2023,
  \url{https://www.intel.com/content/www/us/en/developer/articles/technical/intel-sdm.html}.

\bibitem{xeon8480}
------, \emph{Intel® Xeon® Platinum 8480+ Processor}, 2024,
  \url{https://www.intel.com/content/www/us/en/products/sku/231746/intel-xeon-platinum-8480-processor-105m-cache-2-00-ghz/specifications.html}.

\bibitem{amx-accelerate}
------, \emph{Accelerate AI workloads with Intel AMX}, 2022,
  \url{https://www.intel.com/content/dam/www/central-libraries/us/en/documents/2022-12/accelerate-ai-with-amx-sb.pdf}.

\bibitem{mixedprecision}
\BIBentryALTinterwordspacing
A.~Abdelfattah, H.~Anzt, E.~G. Boman, E.~C. Carson, T.~Cojean, J.~J. Dongarra,
  M.~Gates, T.~Gr{\"{u}}tzmacher, N.~J. Higham, X.~S. Li, N.~Lindquist, Y.~Liu,
  J.~A. Loe, P.~Luszczek, P.~Nayak, S.~Pranesh, S.~Rajamanickam, T.~Ribizel,
  B.~Smith, K.~Swirydowicz, S.~J. Thomas, S.~Tomov, Y.~M. Tsai, I.~Yamazaki,
  and U.~M. Yang, ``A survey of numerical methods utilizing mixed precision
  arithmetic,'' \emph{CoRR}, vol. abs/2007.06674, 2020. [Online]. Available:
  \url{https://arxiv.org/abs/2007.06674}
\BIBentrySTDinterwordspacing

\bibitem{choquette2023nvidia}
J.~Choquette, ``Nvidia hopper h100 gpu: Scaling performance,'' \emph{IEEE
  Micro}, 2023.

\bibitem{jouppi2021ten}
N.~P. Jouppi, D.~H. Yoon, M.~Ashcraft, M.~Gottscho, T.~B. Jablin, G.~Kurian,
  J.~Laudon, S.~Li, P.~Ma, X.~Ma \emph{et~al.}, ``Ten lessons from three
  generations shaped google’s tpuv4i: Industrial product,'' in \emph{2021
  ACM/IEEE 48th Annual International Symposium on Computer Architecture
  (ISCA)}.\hskip 1em plus 0.5em minus 0.4em\relax IEEE, 2021, pp. 1--14.

\bibitem{areaFMA}
J.~Tong, D.~Nagle, and R.~Rutenbar, ``Reducing power by optimizing the
  necessary precision/range of floating-point arithmetic,'' \emph{IEEE
  Transactions on Very Large Scale Integration (VLSI) Systems}, vol.~8, no.~3,
  pp. 273--286, 2000.

\bibitem{area8bit}
N.~Wang, J.~Choi, D.~Brand, C.-Y. Chen, and K.~Gopalakrishnan, ``Training deep
  neural networks with 8-bit floating point numbers,'' in \emph{Proceedings of
  the 32nd International Conference on Neural Information Processing Systems},
  ser. NIPS'18.\hskip 1em plus 0.5em minus 0.4em\relax Red Hook, NY, USA:
  Curran Associates Inc., 2018, p. 7686–7695.

\bibitem{linpack}
J.~J. Dongarra, C.~B. Moler, J.~R. Bunch, and G.~W. Stewart, \emph{LINPACK
  users' guide}.\hskip 1em plus 0.5em minus 0.4em\relax SIAM, 1979.

\bibitem{hpcg}
\BIBentryALTinterwordspacing
M.~A. Heroux, J.~Dongarra, and P.~Luszczek, ``Hpcg benchmark technical
  specification,'' Sandia National Lab.(SNL-NM), Albuquerque, NM (United
  States), Tech. Rep., 10 2013. [Online]. Available:
  \url{https://www.osti.gov/biblio/1113870}
\BIBentrySTDinterwordspacing

\bibitem{yamada2018vector}
Y.~Yamada and S.~Momose, ``{Vector engine processor of NEC’s brand-new
  supercomputer SX-Aurora TSUBASA},'' in \emph{Proceedings of A Symposium on
  High Performance Chips, Hot Chips}, vol.~30, 2018, pp. 19--21.

\bibitem{onednn}
Intel, ``Oneapi deep neural network library,'' 2024,
  \url{https://oneapi-src.github.io/oneDNN/}.

\bibitem{paszke2019pytorch}
A.~Paszke, S.~Gross, F.~Massa, A.~Lerer, J.~Bradbury, G.~Chanan, T.~Killeen,
  Z.~Lin, N.~Gimelshein, L.~Antiga \emph{et~al.}, ``Pytorch: An imperative
  style, high-performance deep learning library,'' \emph{Advances in neural
  information processing systems}, vol.~32, pp. 8026--8037, 2019.

\bibitem{tensorflow2016}
M.~Abadi, P.~Barham, J.~Chen, Z.~Chen, A.~Davis, J.~Dean, M.~Devin,
  S.~Ghemawat, G.~Irving, M.~Isard \emph{et~al.}, ``Tensorflow: A system for
  large-scale machine learning,'' in \emph{{12th USENIX Symposium on Operating
  Systems Design and Implementation (OSDI 16)}}, 2016, pp. 265--283.

\bibitem{xbyak}
M.~Shigeo, ``Xbyak, a c++ jit assembler for x86 (ia32), x64 (amd64, x86-64),''
  2024, \url{https://github.com/herumi/xbyak}.

\bibitem{xbyakArm}
K.~Kawakami, K.~Kurihara, M.~Yamazaki, T.~Honda, and N.~Fukumoto, ``A binary
  translator to accelerate development of deep learning processing library for
  aarch64 cpu,'' \emph{IEICE Transactions on Electronics}, vol. 105, no.~6, pp.
  222--231, 2022.

\bibitem{ferrari2023advancing}
V.~Ferrari, R.~Sousa, M.~Pereira, J.~P. L.~De~Carvalho, J.~N. Amaral,
  J.~Moreira, and G.~Araujo, ``Advancing direct convolution using convolution
  slicing optimization and isa extensions,'' \emph{ACM Transactions on
  Architecture and Code Optimization}, vol.~20, no.~4, pp. 1--26, 2023.

\bibitem{marcel2010torchvision}
S.~Marcel and Y.~Rodriguez, ``Torchvision the machine-vision package of
  torch,'' in \emph{Proceedings of the 18th ACM international conference on
  Multimedia}, 2010, pp. 1485--1488.

\bibitem{benchdnn}
Intel, ``Benchdnn github repository,'' 2024,
  \url{https://github.com/oneapi-src/oneDNN/blob/master/tests/benchdnn/README.md}.

\bibitem{mcpat09}
S.~Li, J.~H. Ahn, R.~D. Strong, J.~B. Brockman, D.~M. Tullsen, and N.~P.
  Jouppi, ``{McPAT: An Integrated Power, Area, and Timing Modeling Framework
  for Multicore and Manycore Architectures},'' in \emph{International Symposium
  on Microarchitecture (MICRO)}, 2009, pp. 469--480.

\bibitem{pcacti}
\BIBentryALTinterwordspacing
``Pcacti,'' 2025. [Online]. Available:
  \url{https://sportlab.usc.edu/downloads/packages}
\BIBentrySTDinterwordspacing

\bibitem{FINCACTI}
A.~Shafaei, Y.~Wang, X.~Lin, and M.~Pedram, ``Fincacti: Architectural analysis
  and modeling of caches with deeply-scaled finfet devices,'' in \emph{2014
  IEEE Computer Society Annual Symposium on VLSI}, 2014, pp. 290--295.

\bibitem{pcacti-5nm}
\BIBentryALTinterwordspacing
Q.~Xie, X.~Lin, Y.~Wang, M.~J. Dousti, A.~Shafaei, M.~Ghasemi{-}Gol, and
  M.~Pedram, ``5nm finfet standard cell library optimization and circuit
  synthesis in near-and super-threshold voltage regimes,'' in \emph{{IEEE}
  Computer Society Annual Symposium on VLSI, {ISVLSI} 2014, Tampa, FL, USA,
  July 9-11, 2014}.\hskip 1em plus 0.5em minus 0.4em\relax {IEEE} Computer
  Society, 2014, pp. 424--429. [Online]. Available:
  \url{https://doi.org/10.1109/ISVLSI.2014.101}
\BIBentrySTDinterwordspacing

\bibitem{Xi.2015.HPCA}
S.~Xi, H.~Jacobson, P.~Bose, G.-Y. Wei, and D.~Brooks, ``Quantifying sources of
  error in {McPAT} and potential impacts on architectural studies,'' in
  \emph{International Symposium on High Performance Computer Architecture
  (HPCA)}, 2015, pp. 577--589.

\bibitem{amx-measure}
Intel, \emph{Tips to Measure the Performance of Matrix Multiplication Using
  Intel® MKL}, 2024,
  \url{https://www.intel.com/content/www/us/en/developer/articles/technical/a-simple-example-to-measure-the-performance-of-an-intel-mkl-function.html}.

\bibitem{torchvision2016}
T.~maintainers and contributors, ``Torchvision: Pytorch's computer vision
  library,'' \url{https://github.com/pytorch/vision}, 2016.

\bibitem{wolf-etal-2020-transformers}
\BIBentryALTinterwordspacing
T.~Wolf, L.~Debut, V.~Sanh, J.~Chaumond, C.~Delangue, A.~Moi, P.~Cistac,
  T.~Rault, R.~Louf, M.~Funtowicz, J.~Davison, S.~Shleifer, P.~von Platen,
  C.~Ma, Y.~Jernite, J.~Plu, C.~Xu, T.~L. Scao, S.~Gugger, M.~Drame, Q.~Lhoest,
  and A.~M. Rush, ``Transformers: State-of-the-art natural language
  processing,'' in \emph{Proceedings of the 2020 Conference on Empirical
  Methods in Natural Language Processing: System Demonstrations}.\hskip 1em
  plus 0.5em minus 0.4em\relax Online: Association for Computational
  Linguistics, Oct. 2020, pp. 38--45. [Online]. Available:
  \url{https://www.aclweb.org/anthology/2020.emnlp-demos.6}
\BIBentrySTDinterwordspacing

\bibitem{fujita2023calculation}
K.~Fujita, T.~Yamaguchi, Y.~Kikuchi, T.~Ichimura, M.~Hori, and L.~Maddegedara,
  ``Calculation of cross-correlation function accelerated by tensorfloat-32
  tensor core operations on nvidia’s ampere and hopper gpus,'' \emph{Journal
  of Computational Science}, vol.~68, p. 101986, 2023.

\bibitem{liu2021optimizing}
J.~Liu, D.~Yang, and J.~Lai, ``Optimizing winograd-based convolution with
  tensor cores,'' in \emph{Proceedings of the 50th International Conference on
  Parallel Processing}, 2021, pp. 1--10.

\bibitem{jorda2019performance}
M.~Jorda, P.~Valero-Lara, and A.~J. Pena, ``Performance evaluation of cudnn
  convolution algorithms on nvidia volta gpus,'' \emph{IEEE Access}, vol.~7,
  pp. 70\,461--70\,473, 2019.

\end{thebibliography}


\end{document}